\documentclass[12pt,preprint]{emulateapj}
\usepackage{lscape,url}
\begin{document}

\title{Three Wide Planetary-Mass Companions to FW Tau, ROXs 12, and ROXs 42B}

\author{
Adam L. Kraus\altaffilmark{1,2}, 
Michael J. Ireland\altaffilmark{3,4}, 
Lucas A. Cieza\altaffilmark{5,6},
Sasha Hinkley\altaffilmark{7},\\
Trent J. Dupuy\altaffilmark{2},
Brendan P. Bowler\altaffilmark{5,7},
Michael C. Liu\altaffilmark{5}
}

\altaffiltext{1}{Department of Astronomy, The University of Texas at Austin, Austin, TX 78712, USA }
\altaffiltext{2}{Harvard-Smithsonian Center for Astrophysics, 60 Garden St, Cambridge, MA 02138, USA}
\altaffiltext{3}{Department of Physics and Astronomy, Macquarie University, NSW 2109, Australia}
\altaffiltext{4}{Australian Astronomical Observatory, PO Box 296, Epping NSW 1710, Australia}
\altaffiltext{5}{Institute for Astronomy, University of Hawaii, 2680 Woodlawn Dr., Honolulu, HI 96822, USA}
\altaffiltext{6}{Universidad Diego Portales, Facultad de Ingenier�a , Av. Ej\'ercito 441, Santiago, Chile}
\altaffiltext{7}{Department of Astronomy, California Institute of Technology, 1200 E. California Blvd, MC 249-17, Pasadena, CA 91125, USA}

\begin{abstract}

We report the discovery of three planetary-mass companions ($M = 6$--20 $M_{Jup}$) in wide orbits ($\rho \sim 150$--300 AU) around the young stars FW Tau (Taurus-Auriga), ROXs 12 (Ophiuchus), and ROXs 42B (Ophiuchus). All three wide planetary-mass companions (``PMCs'') were reported as candidate companions in previous binary survey programs, but then were neglected for $>$10 years. We therefore obtained followup observations which demonstrate that each candidate is comoving with its host star. Based on the absolute $M_{K'}$ magnitudes, we infer masses (from hot-start evolutionary models) and projected separations of $10 \pm 4$ $M_{Jup}$ and $330 \pm 30$ AU for FW Tau b, $16 \pm 4$ $M_{Jup}$ and $210 \pm 20$ AU for ROXs 12 , and $10 \pm 4$ $M_{Jup}$ and $140 \pm 10$ AU for ROXs 42B b. We also present similar observations for ten other candidates which show that they are unassociated field stars, as well as multicolor $JHK'L'$ near-infrared photometry for our new PMCs and for five previously-identified substellar or planetary-mass companions. The NIR photometry for our sample of eight known and new companions generally parallels the properties of free-floating low-mass brown dwarfs in these star-forming regions. However, 5 of the 7 objects with $M < 30 M_{Jup}$ are redder in $K'-L'$ than the distribution of young free-floating counterparts of similar $J-K'$. We speculate that this distinction could indicate a structural difference in circum-planetary disks, perhaps tied to higher disk mass since at least two of the objects in our sample are known to be accreting more vigorously than typical free-floating counterparts.

\end{abstract}


\section{Introduction}

Over the past ten years, direct imaging surveys for extrasolar planets have discovered a small number of planetary-mass companions ($\la$20 $M_{Jup}$; hereafter called PMCs) at $\ga$50 AU orbital radii from young stars in nearby star-forming regions. The prototypical wide PMC, 2M1207-3933 b, consists of a 4--8 $M_{Jup}$ companion located $\rho \sim 40$ AU away from a 35 $M_{Jup}$ brown dwarf \citep[][]{Chauvin:2004bd}. Since its discovery, $\sim$10 other PMCs have been found, most of which orbit higher-mass primaries \citep[$\sim$0.3--2.0 $M_{\odot}$;][]{Neuhauser:2005ea,Luhman:2006lr,Lafreniere:2008oy,Schmidt:2008pd,Ireland:2011fj,Delorme:2013fr,Bowler:2013rt}. Recent exoplanet discoveries like HR 8799 bcde, GJ 504 b, and HD 95086 b \citep[][]{Marois:2008zt,Kuzuhara:2013ly,Rameau:2013gf} suggest that giant planets can form at moderately wide orbital radii ($\sim$50--100 AU). However, there are confirmed PMCs with orbits as wide as 300 AU \citep[i.e., 1RXSJ1609 b and GSC 6214-210 b;][]{Lafreniere:2008oy,Ireland:2011fj}. 

These wide-separation PMCs pose a significant challenge to existing models of star and planet formation. Their orbital radii are so large that it is unlikely that they could form like traditional planetary systems, as the classical core accretion timescale is far too long \citep[$\gg$100 Myr at $a > 100$ AU;][]{Pollack:1996dk} and Class II disks \citep[][]{Andrews:2005qf,Andrews:2007kb} should not become Toomre unstable to direct fragmentation of $\sim$5 $M_{Jup}$ objects at these extreme radii \citep[][and references therein]{Dodson-Robinson:2009uq,Meru:2010rt,Kratter:2011qa}. However, it is equally unlikely that PMCs could form like binary companions during the Class 0/I stages, since they fall near or below the opacity-limited minimum mass \citep[][]{Bate:2005uf}, and should accrete to become stellar or brown dwarf binary companions unless they form at nearly the same time that the circumstellar envelope is exhausted \citep[][]{Kratter:2010yr}. Nonetheless, it appears that one of these outcomes must occur.

We previously conducted a combined analysis of many high-resolution imaging surveys of young stars (Kraus et al. 2013, submitted) which demonstrated that very low-mass companions occur at a frequency of $\ga$1--2\% for orbital radii of 75--750 AU and masses of 2--40 $M_{Jup}$. Given this nontrivial frequency, it appears that they represent a genuine population. However, there are still only $\sim$10 PMCs known in star-forming regions, and only two were discovered in statistically robust surveys \citep[][]{Ireland:2011fj}. Most such PMCs were discovered serendipitously or in unpublished surveys that can not be used for calculating population statistics, so the demographics (frequency, mass function, and semimajor axis distribution) are highly uncertain. Some robust surveys now have been conducted in older populations \citep[][]{Chauvin:2010pt,Vigan:2012eu,Rameau:2012yq,Biller:2013mz,Nielsen:2013zr}, but uncertainties in the ages and dynamical birth environments make their interpretation more complex.

Even in the absence of population statistics, detailed studies of individual PMCs provide important evidence regarding their formation and early evolution. Spectroscopic studies of their atmospheres have shown that PMCs resemble free-floating brown dwarfs with late-M and early-L spectral types \citep[][]{Lafreniere:2008oy,Patience:2010lr,Bonnefoy:2010cq,Bowler:2011kx,Faherty:2013lr,Allers:2013kx}, and hence they do not show evidence for metallicity enhancement or otherwise non-solar bulk compositions \citep[e.g.,][]{Madhusudhan:2011li}. Furthermore, \citet[][]{Bowler:2011kx} demonstrated that GSC 6214-210 b has extraordinarily strong Pa$\beta$ emission, a strong indicator for vigorous accretion from a circumsubstellar disk; the survival of this disk places strong constraints on the past dynamical history, arguing that the PMC formed in situ and was not an ejected out to its current position. Other PMCs also show evidence of possible Pa$\beta$ emission, including GQ Lup b \citep[][]{Seifahrt:2007ve} and CT Cha b \citep[][]{Schmidt:2008pd}, and 1RXSJ1609 might have an unresolved (system) excess at 24 $\mu$m \citep[][]{Bailey:2013fj}. These results indicate that accretion disks could be common; we return to this topic in our discussion of the companion to FW Tau. Finally, partial orbital arcs, such as the existing measurement for GQ Lup b \citep[][]{Neuhauser:2008dp}, also can provide evidence for the formation site and orbital evolution of PMCs.

The discovery of additional PMCs will be crucial in mapping the mass and separation distributions and in enabling more of these detailed studies. In this paper, we present the identification of three planetary-mass companions ($M = 5$--20 $M_{Jup}$) in wide orbits ($\rho \sim 150$--300 AU) around young stars ($\tau \sim 1$--3 Myr). All were identified as candidate binary companions $>$10 years ago, but subsequently ignored in the literature. We also confirm that 10 additional candidates are unassociated field stars, and present multi-color data for our new discoveries and for five previously-identified PMCs. In Section 2, we describe our sample of candidate companions, summarize any previous observations of them, and establish the properties of the host stars. In Section 3, we describe our new observations and how we derived the properties of each candidate. In Section 4, we discuss the detailed properties of three newly-identified companions, summarize the identification of ten candidates as unassociated field stars, and discuss our observations of known PMCs. Finally, in Section 5 we use the combined multi-color data to compare the colors and luminosities of the eight PMCs (3 new and 5 known) to free-floating young BDs and to theoretical models.

\section{The Sample}

\subsection{Candidate Wide Planetary-Mass Companions}

We list the targets considered in this study in Table 1. Five of our targets (ROXs 12, ROXs 42B\footnote{Regarding nomenclature, we note that the ``B'' in ROXs 42B indicates that it is the second-brightest optical counterpart in the error circle for the X-ray source ROXs 42. It does not denote anything regarding the binarity of the system (which is unlikely to be associated with ROXs 42A or ROXs 42C).}, DoAr 22, PDS 70, and FW Tau) were identified as candidate companions during the years 2001-2005, but then were subsequently neglected. \citet[][]{Ratzka:2005nx} conducted a $K$ band speckle imaging survey for stellar binarity in the Ophiuchus star-forming region, and by analyzing their data with shift-and-add stacking, they also identified faint candidate companions to ROXs 12, ROXs 42B A+B, and DoAr 22. During our own observations, we also found an additional faint candidate companion to ROXs 42B that was located interior to the published candidate. \citet[][]{Riaud:2006hc} also discovered a candidate companion to PDS 70, a member of Upper Centaurus-Lupus, while conducting a $K$ band coronagraphic AO imaging survey of young southern stars. Finally, \citet[][]{White:2001jf} discovered a wide candidate companion to FW Tau A+B while obtaining broadband optical colors of known young binary systems in Taurus-Auriga with the Hubble Space Telescope; the candidate was represented by marginal (and red) detections at $R_C$ and $I_C$, plus a significant detection in the narrowband $H\alpha$ filter, indicating that it had significant $H\alpha$ emission.

Another seven of our targets (2M04153916, HD 27659, ScoPMS 42b, GSC 06191-00019, GSC 06793-00994, GSC 06794-00156, and UScoJ1608-1935) were identified to have candidate companions in the course of our own survey programs. Two of these stars (GSC06793-00994 and GSC 06794-00156) were part of the sample of Upper Sco members we described in \citet[][]{Ireland:2011fj}, but their candidate companions could not be tested for association in that work. Another three stars (GSC 06191-00019, ScoPMS 42b, and UScoJ1608-1935) were observed to have faint candidate companions during the same campaign, but we did not include them in our previous analysis because they are binary systems with projected separations that impinge on the PMC semimajor axis range ($\rho > 150$ mas). Finally, two targets (2M04153916 and HD 27659) were identified to have faint candidate companions during a snapshot $K'$ band AO imaging of new Taurus members recently discovered by \citet[][]{Luhman:2009wd} and \citet[][]{Rebull:2010xf}; these observations were primarily intended to screen for wide binary companions in anticipation of future observations with nonredundant aperture masking.

Finally, the last six targets we consider (DH Tau, GQ Lup, 1RXSJ1609, UScoJ1610-1935, GSC 6214-210, and ScoPMS 214) were identified as confirmed or likely low-mass companions to young stars in Taurus \citep[DH Tau;][]{Itoh:2005bs}, Lupus \citep[GQ Lup;][]{Neuhauser:2005ea}, and Upper Sco \citep[the remaining four;][]{Lafreniere:2008oy,Kraus:2009uq,Metchev:2009hh,Ireland:2011fj}. All were shown to be comoving with their neighbor star, and hence are generally assumed to be associated. The only exception is the companion to ScoPMS 214, which \citet[][]{Metchev:2009hh} show to have a spectral type inconsistent with the expected luminosity for a faint companion; the implication is that it is likely to be an unassociated field star with similar motion to Upper Scorpius. We will revisit this identification in Section 4.3.

\subsection{Past Observations}

Many of the candidate companions in our sample were initially (or also) identified by other observing programs, as were several confirmed planetary-mass companions that we are also studying.  We have collated all observations for these targets in Table 2, and will combine them with our own measurements in order to test the association of candidate companions and measure a uniform set of colors for confirmed planetary-mass companions. 

We have adopted most of these observations as stated in their sources, but several seem to require updated parameters. The observation of PDS 70 by \citet[][]{Riaud:2006hc} did not include a position angle, so we have measured its PA from images downloaded from the ESO archive. The observation of FW Tau by \citet[][]{White:2001jf} included a PA which is discrepant by $\sim$111$^o$ from our own measurements, so we also measured an updated value of the PA using the processed images from the Hubble Legacy Archive. In both cases, we conservatively assess an uncertainty of $\sim$0.5$^o$, corresponding to the angle subtended by the FWHM of the companion PSF. Finally, the observations reported by \citet[][]{Ratzka:2005nx} for several companions (including two comoving companions) disagree with other observations by significantly more than their reported uncertainties (which are typically $<$10 mas). Since they reported the detections at $S/N \sim$5 from shift-and-add speckle imaging, then we expect that the minimum positional uncertainties should be $\sim$1/5 $\lambda /D$ or $\sim$30 mas. We therefore have adopted uncertainties of 30 mas for their projected separations and the corresponding subtended angle for their position angles. 

More generally, we note that many observations seem to base their uncertainties on the scatter in their measurements, without considering systematic errors. We therefore also have adopted the systematic uncertainty floors that we assign to our own NIRC2 data (which is typically more stable than most other instruments). Due to uncorrected residuals from geometric distortion \citep[e.g.,][]{Anderson:2003nx,Yelda:2010qf}, all astrometry is given a minimum uncertainty of 2 mas, and due to the uncertain plate scale, all projected separations are given a minimum fractional uncertainty of $10^{-3}$. Finally, due to anisoplanatism and variability of the host stars, all contrast ratios are given a minimum uncertainty of 0.05 mag. These uncertainties could be larger for targets observed using multiple telescopes, instruments, and observing techniques. However, there has never been a global calibration of the platescale of geometric distortion (such as with observations of globular cluster fields) for any significant fraction of high-resolution imaging instruments.

\subsection{Properties of the Host Stars}

We obtained photometric data for the host stars of each candidate from publicly available all-sky surveys. The optical $r'$ magnitudes for most host stars were taken from the 14th Carlsberg Meridianal Catalog \citep[CMC14;][]{Evans:2002sp}. Some targets are too far south or otherwise do not have CMC14 counterparts, so for those stars we adopted the average of the USNO-B1.0 $R$ magnitudes \citep[][]{Monet:2003yt}. Based on the stars which have measurements in both catalogs, the magnitudes are equivalent to within the uncertainties ($R_{USNOB}-r' \sim 0$). The 1.0--2.5 $\mu$m ($J$, $H$, and $K_s$) data were obtained from 2MASS \citep[][]{Cutri:2003jh}. Most of our AO imaging measurements were taken with a $K'$ filter ($\lambda = 2.124 \mu$m), while measurements from the literature often use $K$ ($\lambda = 2.196 \mu$m) or $K_s$ ($\lambda = 2.146 \mu$m), but the conversion terms for late-M and early-L objects are negligible \citep[$\la$0.01-0.02 mag;][]{Carpenter:2001ce} compared to the typical systematic uncertainties of AO photometry. Finally, the 3.0--15 $\mu$m data ($W1$ and $W3$) were obtained from WISE \citep[][]{Wright:2010ij}. The WISE $W1$ filter is centered at 3.35 $\mu$m (2.9--3.8 $\mu$m), while our ground-based measurements were taken with an MKO $L'$ filter centered at 3.78 $\mu$m (3.4--4.1 $\mu$m), so use of WISE data requires a color correction. The conversion has not yet been reported in the literature, so we downloaded WISE data for 40 M3.0-L9.5 field dwarfs that have $L'$ data reported in the literature \citep[][]{Leggett:2010th}, and computed a color-SpT relation of $W1 - L' = 0.044 \times SpT -0.078$ (where SpT=0 for M0 and 20 for T0), which has a scatter about the relation of $\sigma = 0.10$ mag for the full sample and $\sigma = 0.05$ mag specifically for M5-L1 dwarfs. We tested this relation for the isolated brown dwarfs observed with $L'$ photometry by \citet[][]{Jayawardhana:2003hs}, and found that the mean $W1-L'$ colors agreed to within 0.2 magnitudes, with a scatter of 0.2 magnitudes; similar offsets were seen for disk-bearing and disk-free brown dwarfs, indicating that infrared excesses do not bias this relation.

None of the host stars were observed by HIPPARCOS or otherwise have direct trigonometric parallaxes, so we must infer their distances from the mean and standard deviation of other members of their associations. Based on previous trigonometric parallaxes, we have adopted these values collectively for all members of Taurus \citep[145 $\pm$ 15 pc;][]{Torres:2009ct}, Ophiuchus \citep[120 $\pm$ 10 pc;][]{Loinard:2008bs}, Upper Sco \citep[145 $\pm$15 pc;][]{de-Zeeuw:1999xv}, and Upper Centaurus-Lupus \citep[140 $\pm$ 15 pc;][]{de-Zeeuw:1999xv}. The number of HIPPARCOS parallaxes toward the Lupus dark clouds is small, so we instead adopt the extinction-based measurement of 155 $\pm$15 pc reported by \citet[][]{Lombardi:2008od}, which is consistent with the convergent-point distance recently reported by \citet[][]{Galli:2013ul}. Where available, we adopt proper motions for each star that were reported in the UCAC3 catalog \citep[][]{Zacharias:2010lr}. However, some targets are too optically faint for UCAC3, so we instead report proper motions derived from USNO-B1, 2MASS, and SDSS using the methods described in \citet[][]{Kraus:2007mz}.

We have adopted the spectral types for the host stars as reported in the literature. Many of the host stars also have extinction measurements reported in the literature. However, some are missing such measurements, and others appear to be significantly redder than expected for the given spectral type and $A_V$. We therefore have calculated new extinction estimates by comparing each star's observed $r'-J$ or $R-J$ color to that for field dwarfs of the same spectral type \citep[][]{Kraus:2007mz}, converting the color excess to a visual extinction using the extinction relations of \citet[][]{Schlegel:1998yj}. Most of our new measurements agree to within $\Delta A_V \la 1$ or $\Delta A_J \la 0.25$. However, two stars (ROXs 12 and DoAr 22) are significantly more extincted than reported in the literature ($\Delta A_V \sim 3$). 

We estimated the temperatures of the host stars based on their spectral types, combined with the dwarf temperature scale of \citet[][]{Schmidt-Kaler:1992ab} for $\le$M0 stars and the pre-main sequence temperature scale of \citet[][]{Luhman:2003pb} for $>$M0 stars. We estimated their luminosities from their dereddened 2MASS $J$ magnitudes, the inferred distances listed above, and the $J$ band bolometric corrections we previously tabulated in \citet[][]{Kraus:2007mz}. Finally, we used these temperatures and luminosities to place each host star on the HR diagram, where they can be directly compared to the predictions of pre-main sequence evolutionary models. Different models make highly discrepant predictions, but based on past calibrations \citep[as summarized by][]{Hillenbrand:2004bh} and the dynamical masses that we have measured in these regions (e.g., Ireland et al. in prep; Kraus et al. in prep), we have chosen the models of \citet{Baraffe:1998yo} using a convective mixing length of 1 $H_P$. The star HD 27659 falls outside the temperature range spanned by those models, so we instead used the models of \citet[][]{Siess:2000ce}. This said, its proper motion is not consistent with Taurus membership, so an age and mass derived for the distance of Taurus might not be meaningful.

We summarize the properties of these host stars in Table 1.

\section{New Observations and Candidate Properties}

\subsection{High-Resolution Imaging Observations and Data Analysis}

Most of our high-resolution imaging observations were obtained at Keck Observatory using the Keck-II 10m telescope and NIRC2, its facility adaptive optics imager. GSC 06191-00019 was also observed at Palomar Observatory using the Palomar-Hale 200'' telescope and PHARO, its facility adaptive optics imager \citep[][]{Hayward:2001io}. All of the primary stars are brighter than $R=15$, so we observed them with natural guide star adaptive optics (NGSAO) at both Keck and Palomar. 

The observations span a number of observing seasons (years 2007-2013) and were taken by several different observers, so they vary significantly in their details (such as total integration time, dither pattern, and Fowler sampling). Several observations using NIRC2 in $K'$ were also taken using the 600 mas coronagraphic spot, which provides $\Delta K' = 7.17 \pm 0.10$ mag of attenuation for the primary star (as measured from observations of binary stars). All observations used the smallest pixel scales (10 mas/pix with NIRC2 and 25 mas/pix with PHARO). We summarize the salient details for these observations in Table 3 and also refer readers to the Keck Observatory Archive\footnote{\url{http://www2.keck.hawaii.edu/koa/public/koa.php}}, which now hosts all NIRC2 data after an 18 month proprietary period.

The data analysis follows the same prescription as described in \citet[][]{Kraus:2008zr}. To briefly summarize, we measured astrometry and aperture photometry for each source with respect to the purported primary star using the IRAF task DAOPHOT \citep[][]{Stetson:1987ya}; all measurements were conducted using apertures of 0.5, 1.0, and 2.0 $\lambda /D$, and then the optimal aperture was chosen to maximize the significance of the detection (given the competing uncertainties from the primary star's PSF halo and the Poisson noise for the candidate companion). In order to estimate the uncertainties from the data, we analyzed the measurements in individual frames and then combined those measurements to estimate the mean and standard deviation. We then corrected for geometric distortion in the images, accounting for the remaining systematic uncertainty in the plates scales and distortion solutions of PHARO \citep[0.3\%; ][]{Ireland:2011fj} and NIRC2 \citep[0.05\%; ][Cameron 2008]{Ghez:2008my} by adding those terms in quadrature with the observed scatter. Candidates were judged to be comoving if their relative proper motion fell within $<$3$\sigma$ of zero and disagreed with the expected motion of a background star by $>$3$\sigma$.

In Table 3, we list the observations and results for all candidate companions that we observed.

\subsection{Candidate Properties}

Young stars are intrinsically variable due to several phenomena (i.e., spots, accretion, and variable extinction), so the conversion of contrasts into magnitudes and colors is subject to  systematic uncertainties. Furthermore, the precision of adaptive optics photometry is fundamentally limited by anisoplanatism \citep[e.g.,][]{Steinbring:2002lr}, especially for relatively faint targets that were observed with modest strehl ratios. To average down these uncertainties, we have combined all available photometric measurements for each object to compute a weighted mean contrast in each filter, with an additional caveat that no epoch was given an uncertainty better than $\sim$0.05 mag. We then computed apparent magnitudes by adding each contrast measurement to the known magnitude for its parent star (Section 2.3; Table 1), adding the uncertainties in quadrature and also assessing a systematic uncertainty of $\sim$0.05 mag in the brightness of the primary \citep[due to its potential variability at the 2MASS epoch;][]{Carpenter:2002yu}. We report the resulting apparent magnitudes in Table 4.

We measured the relative motion of each object with a weighted linear fit to compute ($\mu_{rel,\alpha}$,$\mu_{rel,\delta}$). Due to the precise and accurate calibration of NIRC2 astrometry \citep[e.g., ][]{Yelda:2010qf}, then any fit with multiple NIRC2 points (including those of the three bona fide companions) is effectively dominated by the NIRC2 measurements. If the companion was not associated, then there also  would be an error term for parallactic motion, but such motions are small compared to the proper motion over timescales of $>$1 year. We report the relative motions and the inferred physical projected separations in Table 4.

For the bona fide companions, the absolute magnitudes of the companions were converted into masses (in $M_{Jup}$) using the hot-start DUSTY models, which are more appropriate than the COND models due to the dusty nature of young low-temperature photospheres \citep[][]{Cruz:2009uh,Allers:2010tc,Bowler:2010vn,Patience:2010lr}. The reported uncertainties are dominated by the primary stars' poorly constrained ages, which leave the masses uncertain by up to 50\%; the photometric uncertainties are negligible in comparison. The plausibility of the models adds another significant systematic uncertainty; newer models suggest that the hot-start models could significantly underestimate planet masses \citep[][; Marleau et al. 2013]{Fortney:2008lo,Spiegel:2012ys}.

In each case, we use the hot-start DUSTY models to estimate a range of plausible masses based on the age range seen in its host region (1--5 Myr for Taurus, Ophiuchus, and Lupus; 5--10 Myr for Upper Scorpius), and assign a best-fit mass which is the average of these minimum and maximum limits. These age ranges encompass a possible upward revision of all pre-main sequence stellar ages \citep[e.g., ][]{Pecaut:2012dp}, and hence allow for higher masses than were previously estimated. We will discuss our estimates for the companion masses in more detail in Section 4, and list the mass ranges in Table 4.

\section{Results}

\subsection{Three Planetary-Mass Companions to Young Stars}

As we summarized in Table 4, three of the candidate companions in our sample are comoving with their host stars. Each is located in a wide orbit ($\rho \ga 150$ AU) and has an apparently planetary mass ($M = 6$--20 $M_{Jup}$). All were previously reported in the literature as candidate binary companions, but have been neglected for the past decade. Based on the two-point correlation function in unbound associations like these regions \citep[][]{Kraus:2008fr}, the probability of chance alignment with an unbound association member within $<$3\arcsec\, is small ($<$1 chance alignment, including stellar-mass companions, per $10^4$ members), and hence these objects represent a population of bound companions near or within the planetary-mass regime.

\subsubsection{FW Tau b}

 \begin{figure*}
 \epsscale{1.0}
 \plottwo{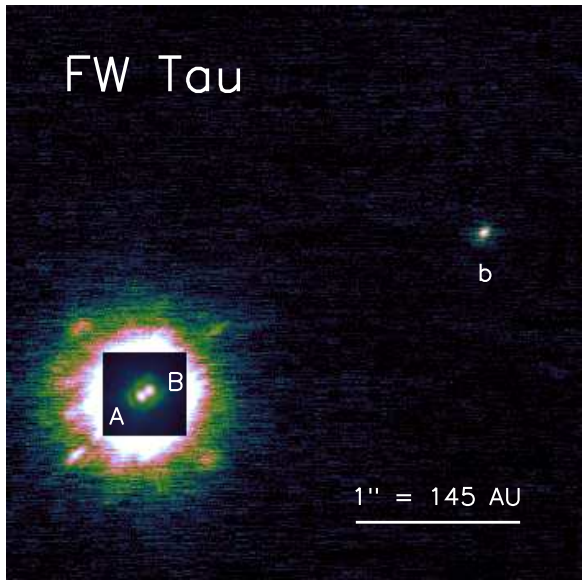}{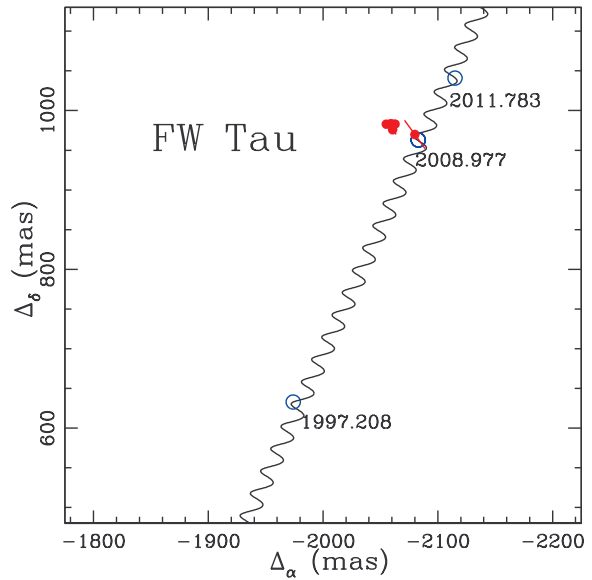}
 \caption{Left: A single $K'$ image of FW Tau from 2008/12/23. The image is not coronagraphic; most of the image is shown with a linear stretch that saturates at 110\% of the peak brightness of the wide companion, while a box of size 0.5\arcsec\, is instead shown with a linear stretch that saturates at 110\% of the peak brightness of the primary star in the close binary. North is up and the angular scale is shown. Right: The relative astrometry ($\Delta_{\alpha}$,$\Delta_{\delta}$) for the companion with respect to the photocenter of the central binary, where the error bars are shown in terms of $\sigma_{\rho}$ and $\sigma_{\theta}$. The solid black line shows the expected track for a nonmoving background star, shifted to minimize the weighted residuals with respect to the expected positions at each epoch of observation (blue circles).}
 \end{figure*}

FW Tau AB is located near the center of the Taurus-Auriga complex and is a close ($\rho \sim$75 mas; 11 AU) binary system comprised of two stars with near-equal fluxes at wavelengths of 0.3--2.2 $\mu$m and equal fluxes at $K'$ band \citep[][]{White:2001jf}. Given the unresolved system spectral type of M4 \citep[][]{Briceno:1993hc}, models suggest individual masses of 0.28 $\pm$0.05 $M_{\odot}$ for each component. The corresponding isochronal age is 1.8$^{+1.0}_{-0.5}$ Myr, consistent with the median age of Taurus-Auriga \citep[1.8 Myr;][]{Kraus:2009fk}. Observations of the $H\alpha$ emission line strength (EW[H$\alpha$]=-17\AA; Brice\~no et al. 1993) and the lack of excess emission from near-IR and mid-IR photometry \citep[][]{Rebull:2010xf} suggest that the primary is not accreting and does not host a substantial disk around either component or within the inner $\sim$50 AU. However, \citet[][]{Andrews:2005qf} reported a 4$\sigma$ detection of submm flux ($F_{\nu} = 4.5 \pm 1.1$ mJy) that could indicate a disk with a very small mass ($M_{disk} \sim$0.2 $M_{Jup}$) located around one or more objects in the system.

The candidate wide companion to FW Tau was first identified by \citet[][]{White:2001jf}, who noted a faint object at a projected separation of $\sim$2.3\arcsec\, ($\sim$330 AU) from the primary star of the central binary. This candidate companion was marginally detected by HST imaging in the F675W ($R_C$) and F814W ($I_C$) filters, as well as having a significant detection in the narrowband F656N ($H\alpha$) filter. They concluded from its apparently significant $H\alpha$ line emission that it could be an accreting companion. However, no further observations of the candidate companion have since been reported. 

We obtained further observations of the FW Tau system in 2008 and 2011, and as we show in Figure 1 (left), the candidate companion has a significant (but quite faint) NIR counterpart. The multi-epoch astrometry shown in Figure 1 (right) demonstrates that the candidate is indeed a comoving companion, with a net relative motion of $1.7 \pm 0.9$ mas/yr ($\sim$1.1 km/s) with respect to the primary of the central pair.  Based on its dereddened absolute magnitude of $M_{K'} = 9.28 \pm 0.24$, we estimate a companion mass of $10 \pm 4$ $M_{Jup}$ for system ages of 1--5 Myr \citep[][]{Chabrier:2000sh}. The extremely red color of the companion (dereddened $J-K' = 1.73 \pm 0.13$) is consistent with this identification, especially given the low line of sight reddening to FW Tau AB ($A_V = 0.4$; Table 1). This very red $J-K'$ color is typically only seen among field dwarfs for the reddest late-L dwarfs, but as was noted by \citet[][]{Allers:2010tc}, young objects often seem to be redder than field counterparts of similar temperature. Theoretical models predict that the companion should have a temperature of $T_{eff} \sim$1900-2100 K, and hence a spectral type of $\sim$L1--L2. Given its bound nature and apparently planetary mass, we hereafter denote the companion as ``FW Tau (AB) b'', or ``FW Tau b'' for simplicity.

The significant H$\alpha$ emission from FW Tau b, combined with the possible presence of a disk somewhere in the system, indicate a high likelihood that it hosts a (circumplanetary?) disk like that of GSC 6214-210 b \citep[][]{Bowler:2011kx}. The presence of this disk also raises some ambiguity regarding the intrinsic luminosity of the companion, as it could be a stellar or brown dwarf companion with an edge-on disk \citep[like HV Tau C or HK Tau B;][]{Stapelfeldt:1998iv,Duchene:2010xt}. However, the morphology argues against this explanation. Stellar edge-on disks show a characteristic central dark lane that separates the light reflecting from the two surfaces of the disk, and we see only a single unresolved point source.

There are two known substellar hosts of edge-on disks, IRAS 04325+2402 C \citep[][]{Hartmann:1999lr,Scholz:2008fk} and 2MASS J04381486+2611399 \citep[][]{Luhman:2007kl}, both with $SpT \sim$ M7.25 . IRAS 04325+2402 C has a mass of $M < 0.1 M_{sun}$, though the NIR spectrum is heavily veiled and a more precise mass determination is difficult. HST images clearly show the characteristic dark lane seen for higher-mass edge-on disks, indicating that at least some substellar edge-on disk systems follow this morphology. 2MASS J04381486+2611399 is an M7.25 brown dwarf ($M \sim 50 M_{Jup}$) that also appears to host an edge-on disk, but is more easily characterized from its scattered light spectrum because its inner disk is cleared, and hence there is no veiling or near-infrared excess. HST images show a bipolar outflow emerging from the brown dwarf, but there is not a clear dark lane. However, it is $>$2 magnitudes brighter than FW Tau b in $K'$. Assuming similar amounts of attenuation in the $K'$ flux and neglecting possible NIR excess in FW Tau b, then the 1 Myr DUSTY models still indicate that FW Tau b would have $M < 15 M_{Jup}$ if seen in a similar geometry. Nonetheless, a dispositive result would require spectroscopic detection of photospheric features that indicate the $T_{eff}$ of the central object.

\subsubsection{ROXs 42B b}

\begin{figure*}
 \epsscale{1.0}
 \plottwo{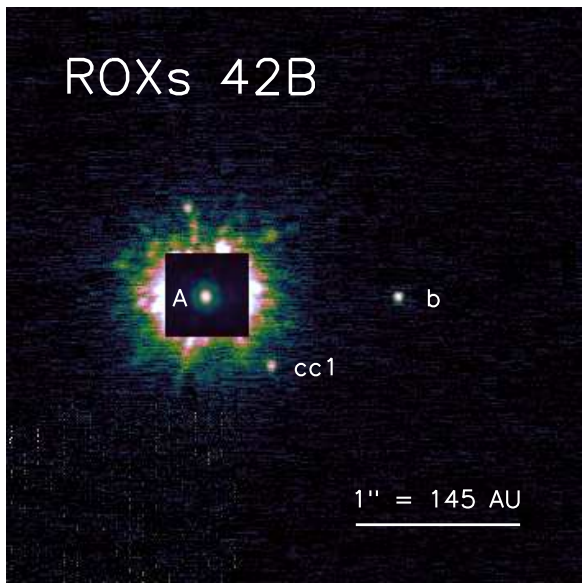}{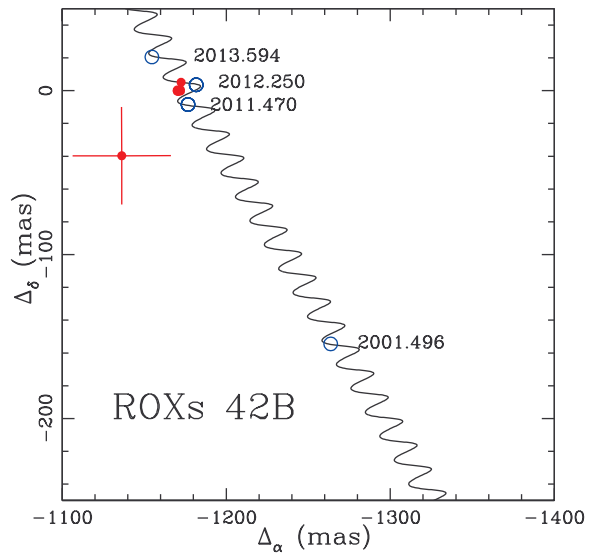}
 \caption{As for Figure 2, but for ROXs 42B b. The image is from 2011/06/22. A long-term superspeckle can be seen above ROXs 42B A; it and other such features can be recognized and rejected by their chromatic behavior in our multi-wavelength data.}
\end{figure*}

\begin{figure}
 \epsscale{1.0}
 \plotone{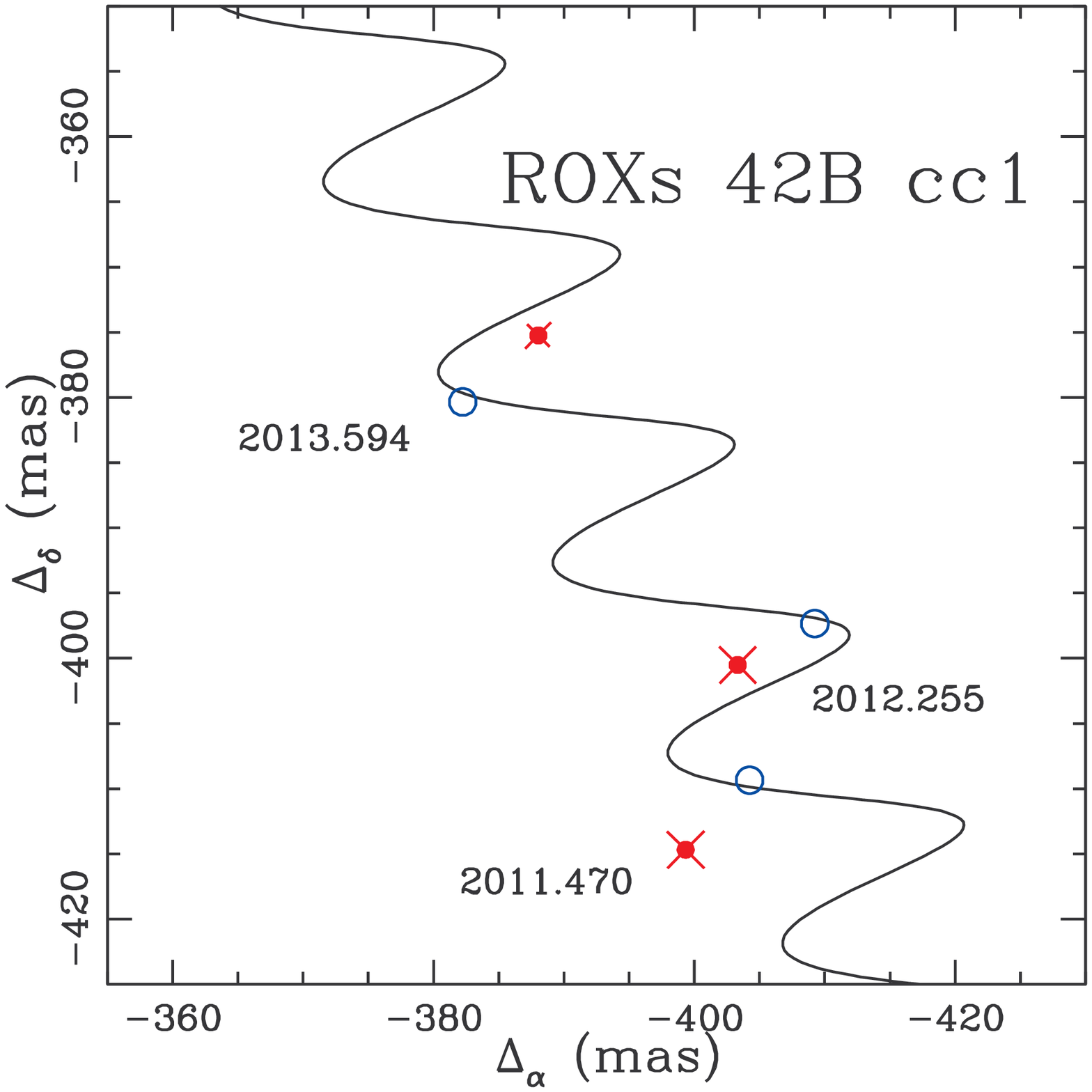}
 \caption{As for the right panel of Figure 3, but for ROXs 42B cc1. The observed motion is small and the probability of finding a background star at $\rho < 1$\arcsec\, is small, but given the broad agreement with the track expected for a nonmoving background star, then the most likely conclusion is that cc1 is not associated with ROXs 42B. The level of disagreement with nonmovement is within the uncertainty in the proper motion of ROXs 42B A ($\pm$3.4 mas/yr).}
\end{figure}

ROXs 42B AB is a little-studied member of the Ophiuchus complex, located $\sim$2$^o$ east of its core. ROXs 42B was identified as a close binary system by \citet[][]{Simon:1995yi} and \citet[][]{Ratzka:2005nx}; the latter survey measured a flux ratio of $\Delta K \sim 1.1$ and a projected separation of $\rho = 83$ mas ($\sim$10 AU). As we will report in a future work, our own observations with nonredundant aperture-mask interferometry have not recovered this companion at a projected separation of $\rho > 15$ mas, suggesting that orbital motion has carried it inward over the past decade. The (unresolved) system was studied with optical spectroscopy by \citet[][]{Bouvier:1992fv}, who determined a spectral type of M0 and identified it as a young star based on the presence of weak H$\alpha$ emission (EW[H$\alpha$] = -1.7\AA), X-ray emission, and possible lithium absorption. Models suggest individual component masses of 0.89$\pm$0.08 $M_{\odot}$ and 0.36$\pm$0.04 $M_{\odot}$, respectively, and a corresponding isochronal age of 6.8$^{+3.4}_{-2.3}$ Myr. The system has no excess in $W3$ and no counterpart in $W4$ from WISE observations, no radio counterpart at 1.3mm \citep[$F_{\nu} < 45$ mJy; $M_{disk} < 9 M_{Jup}$; ][]{Andrews:2007kb}, and H$\alpha$ emission consistent with most WTTSs, and hence there is no evidence that it hosts an optically thick protoplanetary disk.

The binary survey by \citet[][]{Ratzka:2005nx} also identified a faint candidate companion at a much wider projected separation ($\sim$1.1\arcsec; $\sim$140 AU) in the shift-and-added stack of their speckle data. However, it has been neglected in the subsequent literature, so we observed the system in 2011 and 2012 to confirm its existence (Figure 2, left) and test for common proper motion (Figure 2, right). Our observations recovered the candidate companion, but as can be seen in Figure 2 (left), they also revealed another candidate companion even closer to ROXs 42B ($\rho = 0.55$\arcsec; $\sim$70 AU).

As we show in Figure 2 (right), the outer candidate is indeed a comoving companion to ROXs 42B, with a relative motion of $0.7 \pm 1.6$ mas/yr ($\sim$0.5 km/s) with respect to the photocenter of the central binary. Based on its dereddened absolute magnitude of $M_{K'} = 9.39 \pm 0.19$, we estimate a companion mass of $10 \pm 4$ $M_{Jup}$ for system ages of 1--5 Myr \citep[][]{Chabrier:2000sh}. The companion has a luminosity and colors very similar to FW Tau b, so we expect similar atmospheric and bulk properties. Given its bound nature and apparently planetary mass, we hereafter denote the companion (with some regret regarding the nomenclature) as ``ROXs 42B (AB) b'', or ``ROXs 42B b'' for simplicity.

In Figure 3, we show the corresponding proper motion diagram for the inner candidate. The nature of this object is less certain; the proximity to the bright primary leaves its colors and astrometry less reliable, and the larger expected orbital motion ($\sim$6 mas/yr for a circular face-on orbit) is a significant fraction of the absolute proper motion of the system. Nonetheless, the measured astrometry agrees very well with the track expected for a nonmoving background star, with a total relative motion of $17 \pm 3$ mas/yr, and its colors appear to be quite blue ($H - K' = -0.21 \pm 0.11$, $K'-L' = 0.09 \pm 0.14$). We therefore strongly prefer the background hypothesis over the bound companion hypothesis for this object, and we provisionally designate it as a background star (hereafter ``ROXs 42B cc1'').

\subsubsection{ROXs 12 b}

\begin{figure*}
 \epsscale{1.0}
 \plottwo{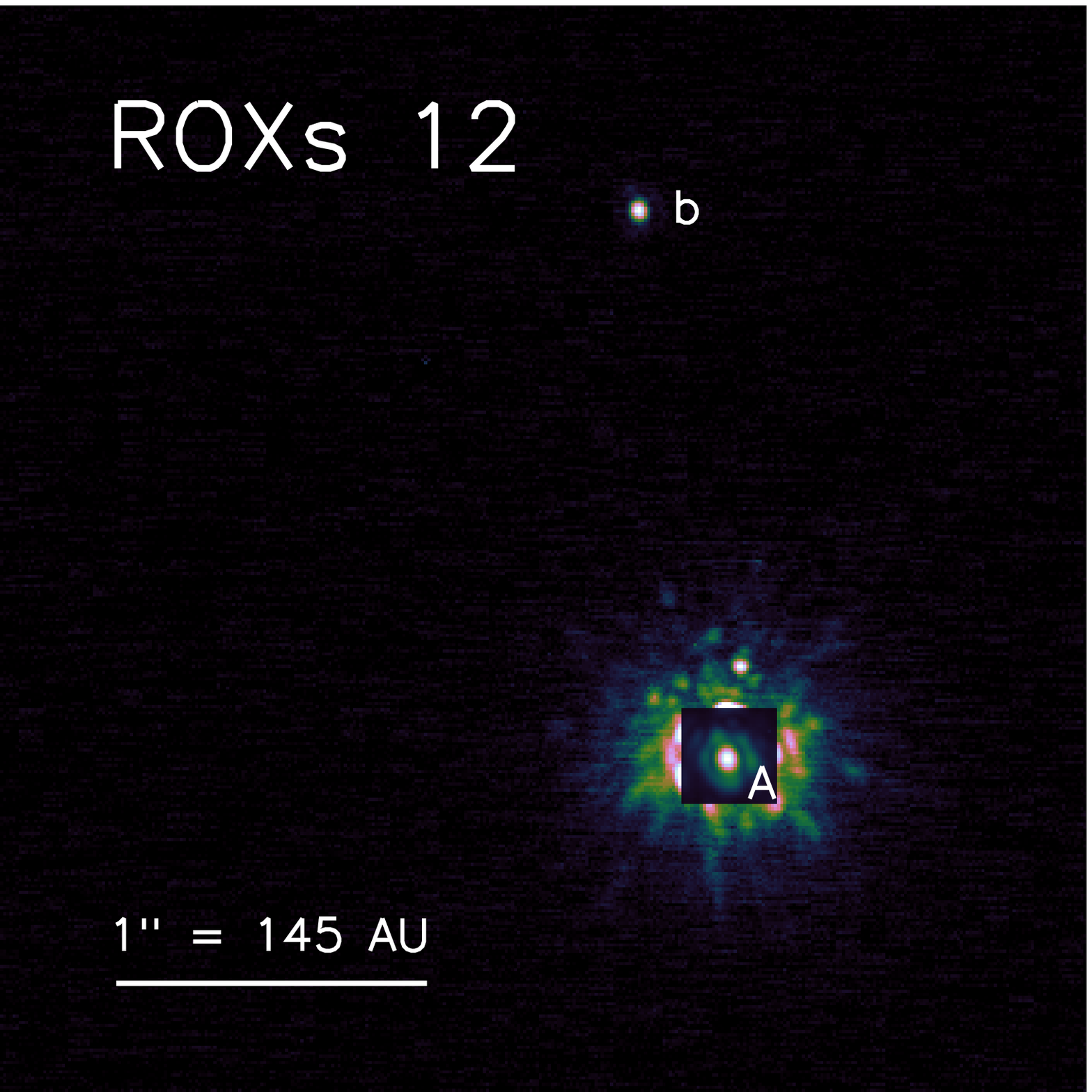}{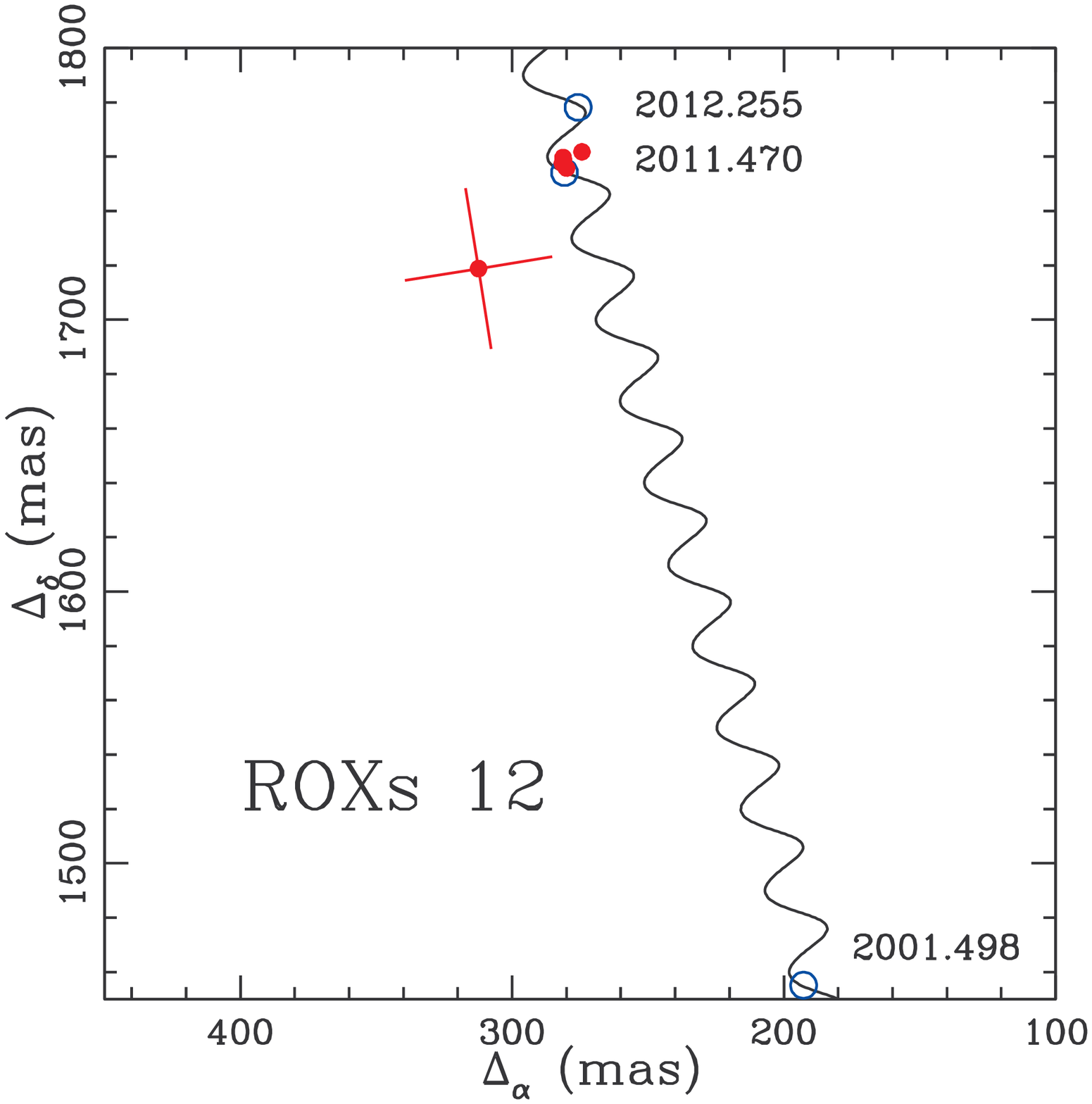}
 \caption{As for Figure 2, but for ROXs 12. The image is from 2011/06/22.}
\end{figure*}

ROXs 12 is another neglected member of the Ophiuchus complex, located $\sim$1$^o$ south of its core. \citet[][]{Bouvier:1992fv} determined a spectral type of M0, and identified it as a young star based on the presence of weak $H\alpha$ emission (EW[H$\alpha$] = -1.2\AA), X-ray emission, and lithium absorption. The HR diagram position of ROXs 12 corresponds to a mass of 0.87$\pm$0.08 $M_{\odot}$ and an age of 7.6$^{+4.1}_{-2.5}$ Myr. A significant excess in the WISE $W3$ and $W4$ filters indicates that the system hosts an optically thick protoplanetary disk, though the nondetection of emission at 1.3mm by \citep[][]{Andrews:2007kb} ($F_{\nu} < 19$ mJy) suggests that the disk is not very massive ($M_{disk} < 4 M_{Jup}$). Also,  the H$\alpha$ line strength is consistent with most WTTSs and much lower than is seen for typical accreting stars.  \citet[][]{Ratzka:2005nx} found no evidence of any close ($\la$1\arcsec) binary companions from speckle interferometry, a result confirmed and extended to much smaller radii by our observations with nonredundant aperture-mask interferometry (Cheetham et al., in prep). \citet[][]{Ratzka:2005nx} did note a faint candidate companion at a projected separation of $\sim$1.7\arcsec\, (210 AU) in the shift-and-added stack of their speckle interferometry data.  However, it subsequently has been neglected in the literature.

We obtained additional observations of ROXs 12 in 2011 and 2012, confirming the existence of the candidate companion (Figure 4, left). The multi-epoch astrometry demonstrates that the candidate companion has a proper motion similar to that of ROXs 12, with a relative motion of $7 \pm 3$ mas/yr ($4.2 \pm 1.6$ km/s) (Figure 4, right). The companion has dereddened $J-H$ and $H-K'$ colors very similar to those of ROXs 42B b, but is significantly redder at longer wavelengths, with $K'-L' = 1.46 \pm 0.11$. While the measurement of comovement is more discrepant than for the other two new PMCs, our results rule out nonmovement at high significance. If the candidate was unassociated, it therefore would be a nearby field star, and the red colors of this candidate clearly distinguish it from the field population. Based on its absolute magnitude of $M_{K'} = 8.42 \pm 0.19$, we estimate a companion mass of $16 \pm 4$ $M_{Jup}$ for system ages of 1--5 Myr \citep[][]{Chabrier:2000sh}. As a bound companion of approximately planetary mass, we denote the companion as ``ROXs 12 b''.

As for FW Tau b, the presence of a disk in the system (determined from the WISE $W3$ and $W4$ photometry) could indicate that we have observed a binary companion which is obscured by an edge-on disk. Given the lack of accretion onto the apparent primary of the system, the argument must be given special consideration for ROXs 12 b. However, as we discussed in Section 4.1.1, the point source morphology seen in our near-infrared images presents the same argument against this interpretation as for FW Tau b. The system is also quite luminous in $W3$ ($m_{W3} = 5.99$), indicating that the companion would need to be massive and intrinsically luminous. The disk therefore is most likely associated with the primary star, and the presence or absence of a disk around the companion can not be inferred from the existing data.

\subsection{Unassociated Background Sources}

\begin{figure*}
 \epsscale{0.8}
 \plotone{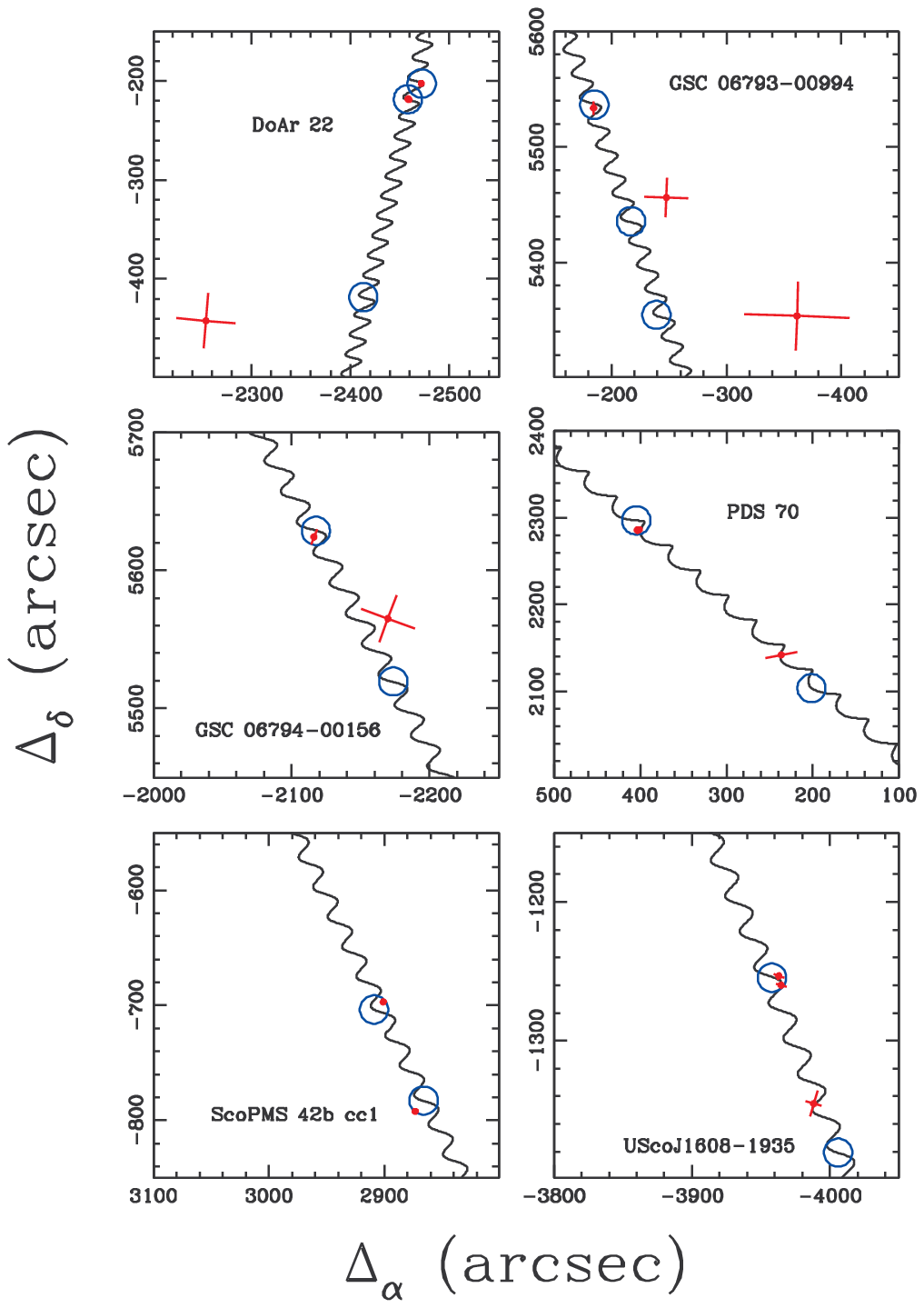}
 \caption{The relative astrometry ($\Delta_{\alpha}$,$\Delta_{\delta}$) for six candidate companions with respect to the photocenter of their candidate host star, where the error bars are show in terms of $\sigma_{\rho}$ and $\sigma_{\theta}$. The solid black line shows the expected track for a nonmoving background star, shifted to minimize the weighted residuals with respect to the expected positions at each epoch of observation (blue circles). In all cases, the track of the candidate companion falls close to the nonmoving track, suggesting that these candidates are unassociated, distant background objects with small proper motions.}
\end{figure*}

\begin{figure*}
 \epsscale{1.0}
 \plotone{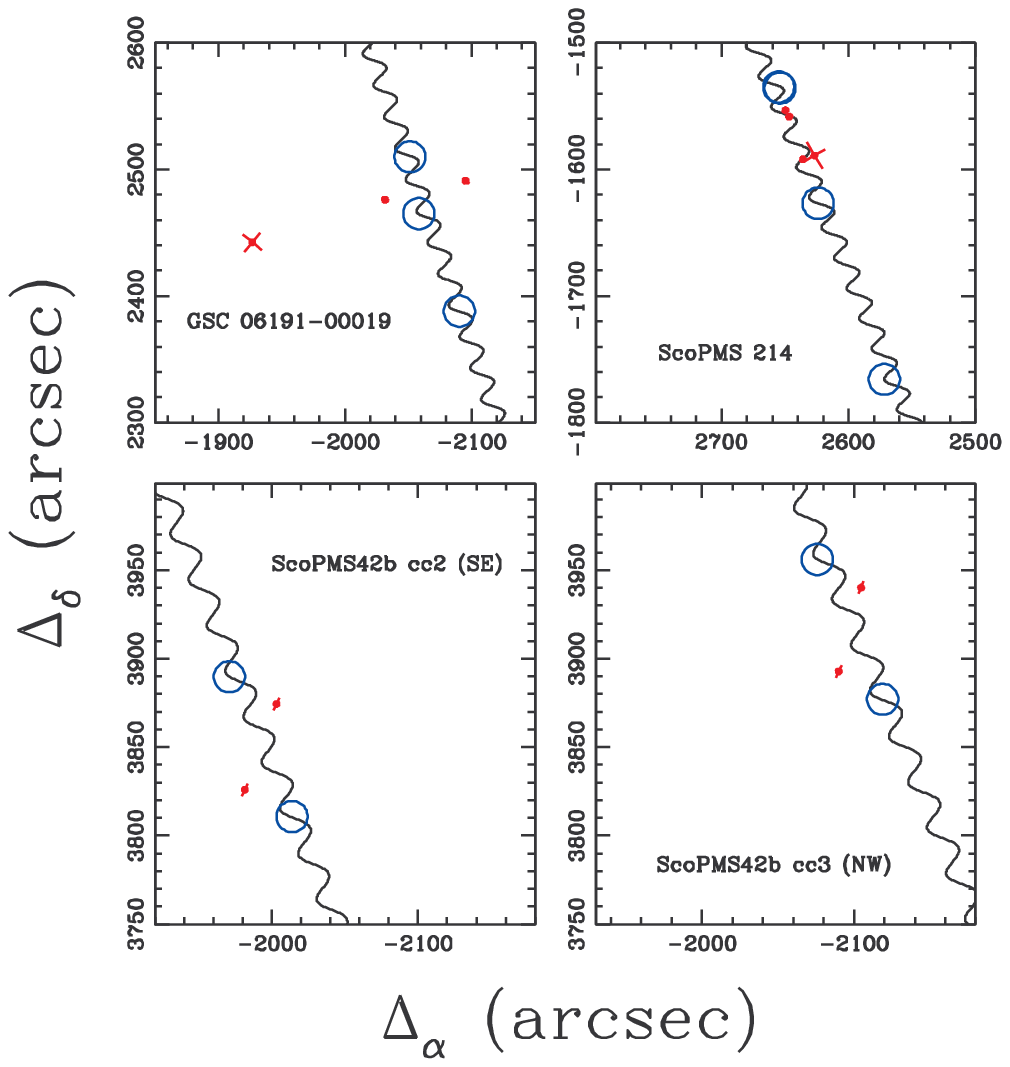}
 \caption{As in Figure 5, but for four candidate companions that display significant motion with respect to either the comoving or nonmoving case.}
\end{figure*}

As we show in Figure 5, six of the candidate companions that we have identified have relative motions that are similar to those expected of a nonmoving background star. We therefore denote these apparent background sources as DoAr 22 cc1, GSC 06793-00994 cc1, GSC 06794-00156 cc1, PDS 70 cc1, UScoJ1608-1935 cc1, and ScoPMS 42b cc1, and have referred to them as such in this work. In most cases, the agreement with comovement is not exact, indicating that the candidate companion has a nonzero absolute proper motion, but these proper motions are generally $\la$10 mas/yr.

In Figure 6, we show the relative motions of four more candidates that appear to be moving with absolute motions of similar magnitude as (but different direction than) their host stars, indicating that they are likely not as distant as the candidates listed above. One target (hereafter denoted GSC 06191-00019 cc1) follows a uniform vector that is perpendicular (and of similar magnitude) to that of its candidate host star. Another target (ScoPMS 214) actually appears to be nearly comoving with its candidate host star, but past spectroscopic observations indicate that it is likely a field dwarf in the solar neighborhood \citep[][]{Metchev:2009hh}. Finally, ScoPMS 42b hosts a pair of faint sources (separated by $\sim$150 mas) that appear to be comoving with each other, but not comoving with the candidate host. These sources were discovered (but not individually resolved) by \citet[][]{Kohler:2000lo}, who found relative astrometry that is also consistent with the tracks that we observe. We infer that these objects (denoted ScoPMS 42b cc2 and ScoPMS 42b cc3) are a binary system comprised of two similar-mass field dwarfs.

Finally, there are two additional targets for which we can not determine a proper motion, but we can classify as unassociated by other criteria. Both candidate companions (2M0415+2813 cc1 and HD 27659 cc1) were identified when we conducted snapshot AO imaging (in preparation for future deep observations) of newly-identified Taurus members identified by \citet[][]{Luhman:2009wd} and \citet[][]{Rebull:2010xf}. We concluded from the snapshot imaging that 2M0415+2813 cc1 was most likely a background galaxy, based on an elliptical elongation of its PSF that is not aligned with the PA to the host star (i.e., anisoplanatism) or with the elevation angle (i.e., windshake or dispersion). Subsequent to this determination, images from SDSS DR9 \citep[][]{Adelman-McCarthy:2011la} detected the candidate in the $g'$ and $r'$ filters with a contrast of 4--5 magnitudes. Since late-M dwarfs have colors of $g'-K' \ga 10$ \citep[][]{Kraus:2007mz}, we conclude the candidate companion is indeed an unassociated background source. The other candidate companion, HD 27659 cc1, has low-quality astrometry from the discovery observations since the candidate fell on a high point in NIRC2's spatially-correlated read noise pattern; even though we have imaging observations from separate seasons, a test of common proper motion is inconclusive. However, the color of the companion is quite blue ($J-L' = 0.12 \pm 0.13$), so it is unlikely to be a planetary-mass companion. The proper motion of HD 27659 is also inconsistent with Taurus membership, so even though its position and distance place it within the Taurus molecular cloud \citep[][]{Kenyon:1994lr}, we suggest that the system is unlikely to be young.

\subsection{Updated Photometry and Astrometry for Known PMCs}

We also have observed five known PMCs in order to obtain more or better colors than were available in the literature, as well as to extend the time baseline in confirming common proper motion. While their identity was not in doubt, we still have analyzed the literature data and our new observations in the same way as for the candidate PMCs (Sections 4.1-4.2). All are once again confirmed to be associated, and we did not find any photometric measurements that contradict measurements in the literature.

GQ Lup continues to show significant evidence of orbital motion in the radial and tangential directions, as first pointed out by \citet[][]{Neuhauser:2008dp}. None of the other targets show clear evidence of motion at more than $\sim 3 \sigma$, which is not a sufficient criterion for significance among such a large sample of measurements. However, the circular orbital velocities at the observed separations ($\sim$100-300 AU) are typically 1--2 km/s or 1--2 mas/yr. We therefore expect that these measurements will achieve statistical significance in the near future, especially with dedicated campaigns that use a single instrument/filter and control or minimize systematic sources of error such as differential chromatic refraction and tilt jitter.

\section{The Colors of Wide PMCs}

\begin{figure}
 \epsscale{1.0}
 \plotone{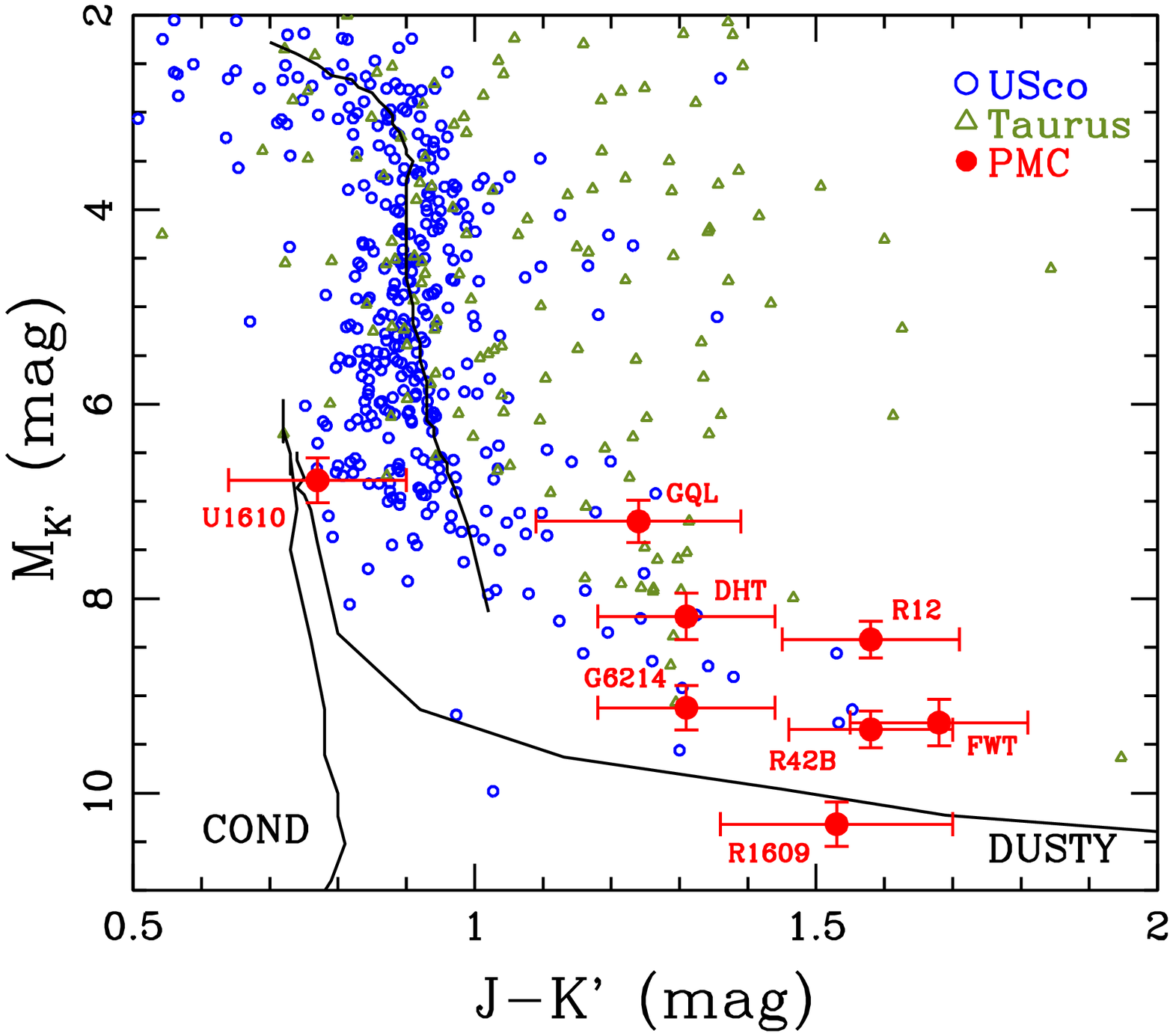}
 \caption{Color-magnitude diagram showing dereddened ($J-K'$,$M_K'$) for our sample of wide PMCs (red filled circles). We also show a representative sample of free-floating members of Upper Sco (blue circles) and Taurus (magenta triangles). The Upper Sco sample was adopted from the census by \citep[][]{Kraus:2007ve}, supplemented by \citet[][]{Slesnick:2008mi} and \citet[][]{Lodieu:2008zg}, and dereddened for an average extinction of $A_V = 1$. The Taurus sample was adopted from \citet[][]{Luhman:2010cr} and \citet[][]{Rebull:2010xf}, dereddened individually with values from the references therein, and screened of binaries as described by \citet[][]{Kraus:2011tg} and \citet[][]{Kraus:2012bh}. Finally, we also show the 5 Myr Lyon evolutionary models of \citet{Baraffe:1998yo}, \citet[][]{Chabrier:2000sh}, and \citet[][]{Baraffe:2003by}. The PMCs follow the Upper Sco sequence in transitioning from the BCAH98 track to the DUSTY track. Once in the DUSTY regime ($M_K > 8$), the sequence appears to decline steeply; the brighter PMCs and Upper Sco members are overluminous for their color, while 1RXSJ1609 is underluminous and not significantly redder, indicating a possible transition toward the COND track. }
\end{figure}

\begin{figure}
 \epsscale{1.0}
 \plotone{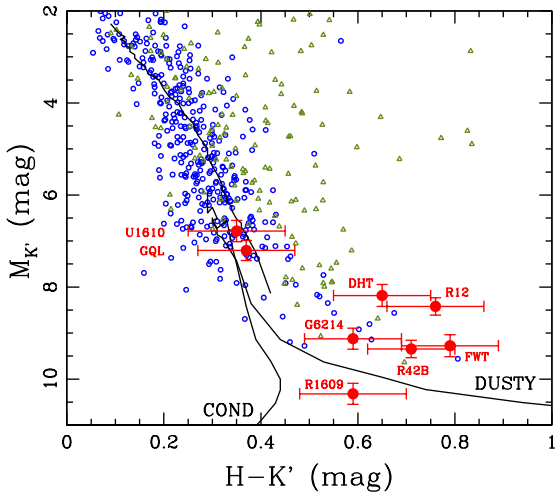}
 \caption{As for Figure 7, but showing dereddened ($H-K'$,$M_K'$). As in Figure 8, most of the PMCs in the DUSTY regime are overluminous, while 1RXSJ1609 is underluminous and might be trending toward the COND (dust-free) sequence.}
\end{figure}

\begin{figure}
 \epsscale{1.0}
 \plotone{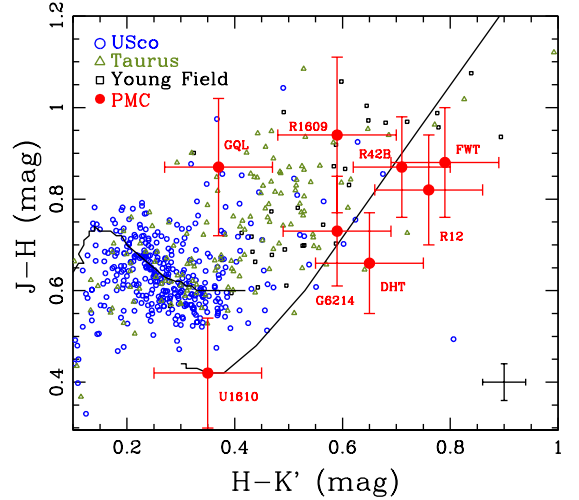}
 \caption{Color-color diagram showing dereddened ($H-K'$,$J-H$) for our sample of wide PMCs (red filled circles). We also show the reference populations described in Figure 7 (Upper Sco in blue and Taurus in magenta), as well as a sample of young field M/L dwarfs \citep[black squares; ][]{Allers:2013kx} that is meant to extend the young sequence past the current limits of star-forming regions. We also show the 5 Myr Lyon models of \citet{Baraffe:1998yo} and \citet[][]{Chabrier:2000sh}. The PMCs and Upper Sco members follow the same ($H-K'$,$J-H$) track as the models, though several PMCs are marginally redder in $H-K'$ than the young sequence.}
\end{figure}

\begin{figure}
 \epsscale{1.0}
 \plotone{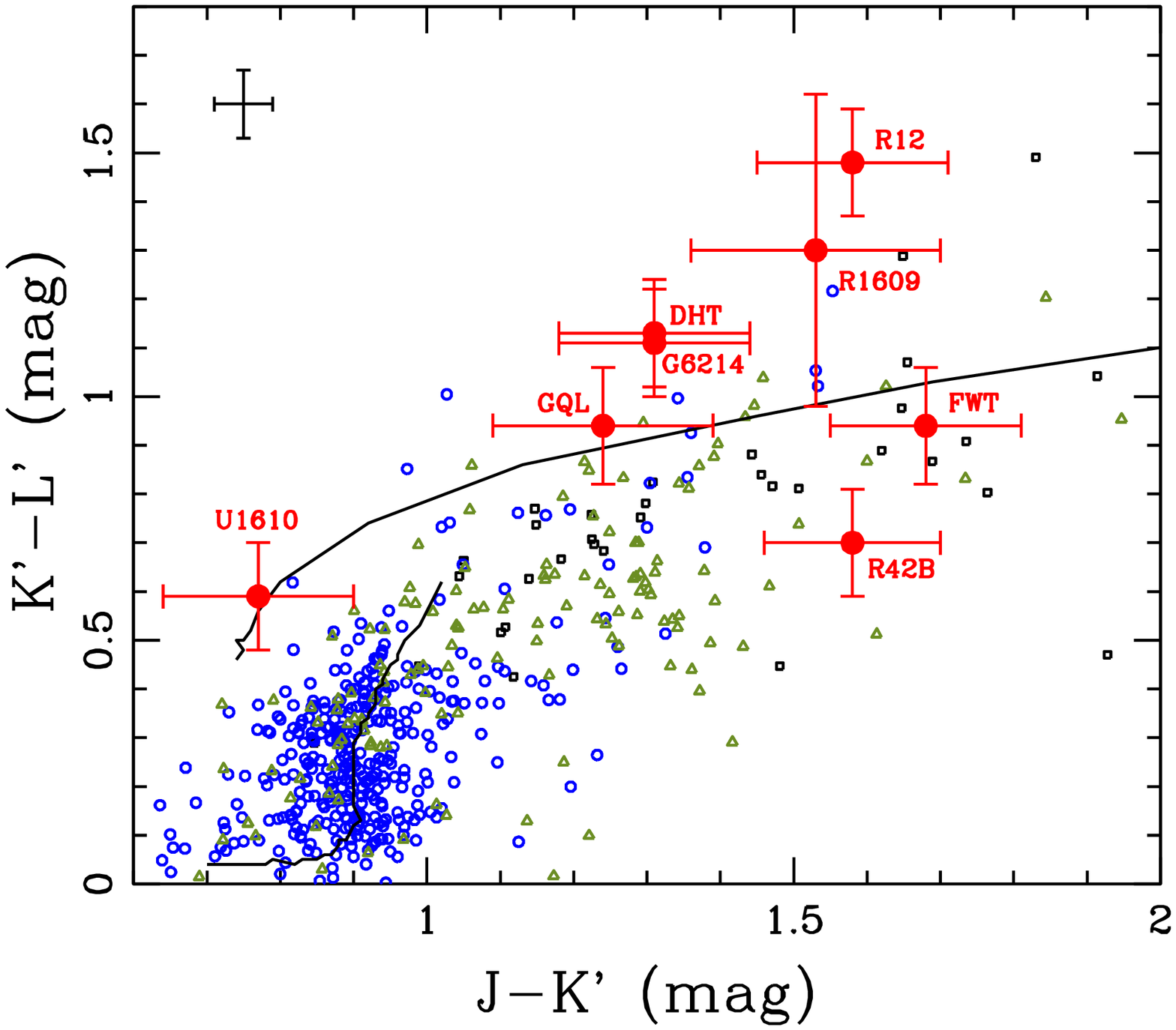}
 \caption{As for Figure 9, but showing dereddened ($J-K'$,$K'-L'$). Unlike for the marginal detection of an $H-K'$ excess in Figure 9, there is a clear excess in $K'-L'$ for many PMCs compared to all of the reference populations of free-floating young objects.}
 \end{figure}

Observations of free-floating young BDs with mid-M to mid-L spectral types show that they follow distinct color-SpT sequences from their older (and more massive) main-sequence counterparts. In particular, young BDs show systematically redder $J-K'$ and $K'-L'$ colors \citep[e.g.,][]{Allers:2010tc}, trends which are often attributed to larger dust abundances or low surface gravities. If PMCs follow the same trend, then it would indicate a similar bulk composition and hence possibly a similar formation mechanism. Significant differences in colors could indicate either a different composition (i.e., from formation in a chemically differentiated disk) or the presence of circum(sub)stellar phenomena like disks, outflows, and accretion.

In Figures 7 and 8, we show ($K'$,$J-K'$) and ($K'$,$H-K'$) color-magnitude diagrams for our sample of PMCs, as well as reference populations of dereddened objects in Upper Sco ($\tau = 5$--10 Myr) and Taurus ($\tau = 1$--5 Myr). We also show the predictions of the 5 Myr BCAH98, DUSTY, and COND model tracks. In both cases, the PMCs in our sample are consistent with the color/luminosity sequence and intrinsic spread for the young sequence. The young sequence seems to transition from the BCAH98 models to the DUSTY models at $J-K \sim 1.1-1.3$, as do the PMCs. However, there is only a modest level of agreement with the DUSTY models at $M_K \ga 8$. At $8 < M_K < 10$, the young sequence and most of the PMCs sit well above the 5 Myr DUSTY track, implying that the models underestimate luminosities in this regime. However, the faintest PMC (1RXSJ1609) has nearly the same color as more luminous objects and falls below the DUSTY track. This object could represent the point where the $J-K'$ color saturates at a maximum value \citep[e.g.,][]{Allers:2010tc} and begins to transition toward the COND tracks.

In general, the PMCs from younger regions (Taurus, Ophiuchus, and Lupus) sit higher on the CMD than those from Upper Sco, indicating that part of the scatter in observed PMC properties might be due to evolution. However, the properties of the primary stars do not always follow this trend. The isochronal ages for ROXs 12 and ROXs 42B are similar to those of GSC 6214-210 and 1RXSJ1609, even though they are nominally members of a younger population. This discrepancy could mean that the spectral types from the literature (which typically date to $\la$1990) need to be repeated with better instruments or reclassified with modern spectral classification systems.

In Figure 9, we show the ($J-H$,$H-K'$) color-color diagram for our sample. Following the format of Figures 7 and 8, we show the BCAH98 and DUSTY models and our reference populations; we also add a sample of young M/L field dwarfs \citep[][]{Allers:2013kx} to illustrate the likely extension of the Taurus and Upper Sco sequences, which have no members later than L2. As for the CMDs, the PMCs largely fall along the young sequence. However, several objects fall redder than the young sequence in $H-K'$, suggesting a possible difference in their properties.

Finally, in Figure 10 we show a corresponding ($J-K'$,$K'-L$) color-color diagram for our sample. Past observations of these objects have found very red $K'-L'$ colors, significantly exceeding those seen for main-sequence field dwarfs. Intriguingly, we find that five of the seven PMCs with $M < 30 M_{Jup}$ are redder than even the young sequence. \citet[][]{Ireland:2011fj} interpreted the potentially red colors of GSC 6214-210 as a sign that it could host an optically thick disk; the existence of such a disk was confirmed by the discovery of ongoing accretion by \citet[][]{Bowler:2011kx}. However, it is unclear whether 5/7 PMCs should be expected to host such a disk, and observations of free-floating BDs of similar temperature and luminosity \citep[][]{Luhman:2010cr} show that the excesses from optically thick disks are only evident at $\ga$5 $\mu$m. Our reference population for Taurus includes many disk-bearing brown dwarfs  \citep[][]{Luhman:2010cr,Rebull:2010xf}, suggesting that the disks would need to be structurally different (i.e., with larger inner rim structures) to produce the observed colors.

We therefore suggest either that these objects host disks which are structurally different from those of free-floating brown dwarfs, or that the red $K'-L'$ colors result from different atmospheric properties (i.e., dustier atmospheres or enhanced $H_2O$ absorption) than those of the free-floating population. The former explanation seems very plausible; since these PMCs generally have formed in the immediate neighborhood of a solar-type star, then they could have accreted a significantly more massive disk (perhaps approaching its Toomre stability limit) than a free-floating BD with a relatively anemic envelope. It is unlikely that these red colors denote the presence of lower-mass companions. Pairing of equal-mass (and equal-color) companions would only affect their positions in color-magnitude diagrams, and based on the color-magnitude sequence of the DUSTY models, then pairing of unequal-mass companions should move objects parallel to the color-color sequences, not perpendicular to them.

Further study at mid-IR and submm/mm wavelengths (to characterize the disk) and optical wavelengths (to characterize the accretion) should provide additional insight into the structure and evolution of these circumsubstellar (or circumplanetary) disks. Detailed analysis of the atmospheric properties ($T_{eff}$, $\log(g)$, and $[Fe/H$]) will require spectroscopic followup \citep[e.g.,][]{Lafreniere:2008oy,Bowler:2011kx} to directly compare the atomic lines and molecular bands to those of free-floating counterparts.

\section{Summary}

We have discovered three planetary-mass companions in wide orbits around the young stars FW Tau (in Taurus) and ROXs 12 and ROXs 42B (in Ophiuchus). All three PMCs were reported as candidate companions in previous binary survey programs, but then were neglected for $>$10 years. We demonstrate with our own followup observations that each candidate is comoving with its host star. Based on the absolute $M_{K'}$ magnitudes, we infer masses and projected physical separations of $10 \pm 4$ $M_{Jup}$ and $330 \pm 30$ AU for FW Tau b, $16 \pm 4$ $M_{Jup}$ and $210 \pm 20$ AU for ROXs 12, and $10 \pm 4$ $M_{Jup}$ and $140 \pm 10$ AU for ROXs 42B b. We also have identified ten other candidates to be unassociated field stars. Finally, we have obtained multicolor $JHK'L'$ near-infrared photometry for our three new companions and for five previously-identified companions. The NIR photometry for our sample of eight known and new PMCs generally parallels the properties of free-floating low-mass brown dwarfs in these star-forming regions. However, 5 of the 7 low-mass PMCs with $L'$ photometry are redder in $K'-L'$ than free-floating counters of similar $J-K'$. We speculate that this distinction could indicate a structural difference in the circum-planetary disks, perhaps tied to higher disk mass since at least two of the objects in our sample are known to be accreting vigorously.

\acknowledgements

We thank the referee for providing a thorough and helpful critique of this paper. ALK was supported in part by a Clay Fellowship and by NASA through Hubble Fellowship grant 51257.01 awarded by STScI, which is operated by AURA, Inc., for NASA, under contract NAS 5-26555. MJI acknowledges support from the Australian Research Council's Discovery program (DP1094977) and a Macquarie University Research Development Grant. TD was supported by NASA through Hubble Fellowship grant 51271.01 awarded by STScI, which is operated by AURA, Inc., for NASA, under contract NAS 5-26555. SH and LC were supported by NASA through the Sagan Fellowship program, and LC was supported by theÊALMA-CONICYT fund allocated to the project number 31120009. ML acknowledges support from NSF grant AST09-09222. Finally, this work was supported by a NASA Keck PI Data Award, administered by the NASA Exoplanet Science Institute. Some of the data presented herein were obtained at the W. M. Keck Observatory from telescope time allocated to NASA through the agency's scientific partnership with Caltech and the University of California. The Observatory was made possible by the generous financial support of the W. M. Keck Foundation.

The authors wish to recognize and acknowledge the very significant cultural role and reverence that the summit of Mauna Kea has always had within the indigenous Hawaiian community. We are most fortunate to have the opportunity to conduct observations from this mountain.

Note added in proof: The comovement of ROXs42B b was also confirmed by Currie et al. (2014) based on an independent analysis of archival data, including the 2011 observations we describe in Section 3.

\bibliographystyle{../apj.bst}
\bibliography{../krausbib}

\clearpage
\begin{landscape}

\begin{deluxetable}{llrrrrrrrllrlllll}
 \tabletypesize{\tiny}
 \tablewidth{0pt}
 \tablecaption{Host Stars for Candidate or Confirmed Planetary-Mass Companions}
 \tablehead{
 \colhead{Name} & \colhead{2MASS J} & \colhead{$m_{r'}$} & \colhead{$m_{J}$} & \colhead{$m_{H}$} & \colhead{$m_{Ks}$} & \colhead{$m_{W1}$}& \colhead{$m_{W3}$} & \colhead{$\mu$} & \colhead{SpT} & \colhead{$A_V$} & \colhead{Dist} & \colhead{$T_{eff}$} & \colhead{$M_{bol}$} & \colhead{$\log(\tau)$} & \colhead{$M$} & \colhead{Refs}
 \\
 \colhead{} & \colhead{} & \multicolumn{6}{c}{(mag)} & \colhead{(mas/yr)} & \colhead{} & \colhead{(mag)} & \colhead{(pc)} & \colhead{(K)} & \colhead{(mag)} & \colhead{((Myr)} & \colhead{($M_{\odot}$)}
  }
 \startdata

\multicolumn{5}{l}{\it New Companion Hosts}\\
FW Tau\tablenotemark{a}&04292971+2616532&15.31&10.34&9.68&9.39&9.19&8.79&(8.9,-28.1)$\pm$3.3&M4&2.1&145&3270$^{+70}_{-70}$&6.52$\pm$0.26&6.26$\pm$0.18&0.28$\pm$0.05&1\\
ROXs 12&16262774-2527247&15.80&11.02&9.93&9.21&8.43&5.99&(-9.5,-30.0)$\pm$3.4&M0&4.5&120&3850$^{+100}_{-70}$&5.94$\pm$0.26&6.88$\pm$0.19&0.87$\pm$0.08&2\\
ROXs 42B\tablenotemark{a}&16311501-2432436&13.58&9.91&9.02&8.67&8.45&8.23&(-8.8,-14.6)$\pm$3.0&M0&2.4&120&3850$^{+100}_{-70}$&5.74$\pm$0.26&6.83$\pm$0.18&0.89$\pm$0.08&2\\

\multicolumn{5}{l}{\it Known Companion Hosts}\\
DH Tau&04294155+2632582&13.16&9.77&8.82&8.18&7.11&5.69&(7.1,-17.9)$\pm$3.4&M1&1.4&145&3705$^{+70}_{-70}$&5.21$\pm$0.26&6.21$\pm$0.17&0.82$\pm$0.09&3\\
GQ Lup&15491210-3539051&11.22&8.61&7.70&7.10&6.07&4.34&(-15.1,-23.4)$\pm$2.7&K7&1.5&155&4060$^{+150}_{-100}$&3.67$\pm$0.26&6.09$\pm$0.12&1.37$\pm$0.10&4\\
1RXS J1609&16093030-2104589&12.11&9.82&9.12&8.92&8.77&8.72&(-11.2,-21.9)$\pm$1.5&K7&0.9&145&4060$^{+150}_{-100}$&5.19$\pm$0.27&6.79$\pm$0.21&1.08$\pm$0.08&5\\
UScoJ1610-1913\tablenotemark{a}&16103196-1913062&12.84&10.03&9.26&8.99&8.70&8.53&(-9.8,-22.5)$\pm$1.9&K7&1.9&145&4060$^{+150}_{-100}$&5.12$\pm$0.27&6.76$\pm$0.20&1.09$\pm$0.08&6\\
GSC 6214-210&16215466-2043091&11.94&10.00&9.34&9.15&9.08&8.96&(-18.6,-32.2)$\pm$1.7&K7&0.2&145&4060$^{+150}_{-100}$&5.57$\pm$0.27&6.99$\pm$0.20&1.01$\pm$0.07&7\\

\multicolumn{5}{l}{\it Neighbors of Background Stars}\\
2M0415+2818&04153916+2818586&15.34&10.55&9.61&9.24&8.75&6.89&(11.6,-26.0)$\pm$3.6&M3.75&2.0&130&3305$^{+70}_{-70}$&6.24$\pm$0.26&6.17$\pm$0.15&0.34$\pm$0.06&8\\
HD 27659\tablenotemark{b}&04225462+2823540&8.28&...&...&7.25&7.33&6.78&(-21.3,-23.3)$\pm$7.6&A4&1.1&130&8500$^{+1000}_{-1000}$&1.52$\pm$0.26&6.96$\pm$0.20&2.14$\pm$0.20&8\\
PDS 70&14081015-4123525&11.15&9.55&8.82&8.54&8.03&5.72&(-32.4,-28.5)$\pm$1.7&K5&0.0&140&4350$^{+120}_{-150}$&5.30$\pm$0.26&7.08$\pm$0.20&1.15$\pm$0.09&9\\
GSC 06191-00019\tablenotemark{a}&15590208-1844142&11.51&9.00&8.34&8.11&7.93&7.75&(-8.5,-24.8)$\pm$1.9&K6&1.4&145&4350$^{+120}_{-150}$&4.70$\pm$0.26&6.78$\pm$0.21&1.32$\pm$0.10&7\\
UScoJ1608-1935&16082387-1935518&13.57&10.20&9.47&9.25&9.07&8.91&(-10.5,-25.7)$\pm$2.3&M1&1.4&145&3705$^{+70}_{-70}$&5.65$\pm$0.26&6.57$\pm$0.18&0.75$\pm$0.08&7\\
ScoPMS 42b&16102174-1904067&14.23&10.68&9.91&9.62&9.31&9.09&(-12.5,-25.8)$\pm$3.1&M1&1.7&145&3705$^{+70}_{-70}$&6.04$\pm$0.27&6.76$\pm$0.19&0.72$\pm$0.08&7\\
GSC 06793-00994&16140211-2301021&11.07&9.38&8.77&8.61&8.41&8.35&(-7.1,-20.7)$\pm$0.8&G4&1.3&145&5840$^{+35}_{-35}$&4.23$\pm$0.27&7.72$\pm$0.19&1.20$\pm$0.04&7\\
GSC 06794-00156\tablenotemark{a}&16245136-2239325&9.40&7.78&7.28&7.08&6.99&7.00&(-11.8,-22.1)$\pm$1.4&G6&1.0&145&5700$^{+35}_{-35}$&3.27$\pm$0.27&7.24$\pm$0.07&1.35$\pm$0.04&7\\
DoAr 22&16261932-2343205&12.18&9.73&9.25&9.02&8.82&8.62&(4.5,-20.0)$\pm$1.3&F5&3.1&120&6440$^{+800}_{-400}$&4.23$\pm$0.26&7.89$\pm$0.34&1.24$\pm$0.05&2\\
ScoPMS 214&16294869-2152118&10.92&8.68&8.00&7.76&7.61&7.52&(-9.9,-23.9)$\pm$1.2&K0&1.9&145&5250$^{+170}_{-170}$&3.55$\pm$0.27&7.02$\pm$0.19&1.40$\pm$0.05&10\\
 \enddata
 \tablenotetext{a}{The primary is itself a known binary; the observed flux ratio has been used to correct the $J$ magnitude before computing $M_{bol}$ and placing the primary of the system on the HR diagram. Binary properties were adopted from \citet[][]{Kraus:2011tg} for FW Tau, \citet[][]{Ratzka:2005nx} for ROXs 42B, and \citet[][]{Kraus:2008zr} for UScoJ1610-1913, GSC 06191-00019, and GSC 06794-00156.}
 \tablenotetext{b}{The $J$ and $H$ magnitudes for HD 27659 are flagged as erroneous in 2MASS, so we adopt an assumed color of $J-K=0.05$ \citep[][]{Kraus:2007mz} and the $K_s$ magnitude from 2MASS in order to calculate $M_{bol}$.}
\tablerefs{
1) \citep[][]{White:2001jf},
2) \citep[][]{Ratzka:2005nx},
3) \citep[][]{Itoh:2005bs},
4) \citep[][]{Neuhauser:2005ea},
5) \citep[][]{Lafreniere:2008oy},
6) \citep[][]{Kraus:2007ve},
7) \citet[][]{Ireland:2011fj},
8) This Work,
9) \citep[][]{Riaud:2006hc},
10) \citep[][]{Metchev:2009hh}.
}
 \end{deluxetable}

 \clearpage
\end{landscape}

  \begin{deluxetable*}{lccrrrl}
 \tabletypesize{\tiny}
 \tablewidth{0pt}
 \tablecaption{Literature Measurements of Known and Candidate Planetary-Mass Companions}
 \tablehead{
 \colhead{Name} & \colhead{JD} & \colhead{Filter} & \colhead{$\rho$} & \colhead{PA} & \colhead{$\Delta m$} & \colhead{Ref}
 \\
 \colhead{} & \colhead{(-2400000)} & \colhead{} & \colhead{(mas)} & \colhead{(deg)} & \colhead{(mag)}
 }
 \startdata
1RXS J1609&54584&$K_s$&2215$\pm$6&27.75$\pm$0.10&7.25$\pm$0.18&1\\
1RXS J1609&54635&$K'$&2210$\pm$2&27.62$\pm$0.05&7.27$\pm$0.05&2\\
1RXS J1609&54639&$H$&2222$\pm$6&27.76$\pm$0.10&7.75$\pm$0.07&1\\
1RXS J1609&54639&$J$&2219$\pm$6&27.76$\pm$0.10&8.08$\pm$0.12&1\\
1RXS J1609\tablenotemark{a}&54928&$K'$&2222$\pm$6&27.65$\pm$0.10&...&1\\
1RXS J1609&54983&$K'$&2211$\pm$2&27.61$\pm$0.05&7.23$\pm$0.05&2\\
1RXS J1609\tablenotemark{a}&55014&$K'$&2219$\pm$6&27.74$\pm$0.10&...&1\\
1RXS J1609\tablenotemark{b}&55021&$L'$&...&...&6.1$\pm$0.3&1\\
DH Tau\tablenotemark{a}&51196&$R_CI_C$&2351$\pm$2&139.36$\pm$0.10&...&3\\
DH Tau\tablenotemark{a}&52602&$H$&2340$\pm$6&139.56$\pm$0.17&...&3\\
DH Tau&53013&$K$&2344$\pm$3&139.83$\pm$0.06&6.01$\pm$0.05&3\\
DH Tau\tablenotemark{b}&53013&$H$&...&...&6.14$\pm$0.05&3\\
DH Tau\tablenotemark{b}&53013&$J$&...&...&5.94$\pm$0.05&3\\
DoAr 22&52093&$K$&2297$\pm$30&258.9$\pm$0.7&5.75$\pm$0.11&4\\
FW Tau\tablenotemark{a}&50526&$R_CI_CH\alpha$&2295$\pm$3&295.0$\pm$0.5&...&5\\
GQ Lup&49445&$K$&714$\pm$36&275.5$\pm$1.1&6.24$\pm$0.13&6\\
GQ Lup\tablenotemark{a}&51279&$R_CI_C$&739$\pm$11&275.62$\pm$0.86&...&7\\
GQ Lup\tablenotemark{b}&52473&$K$&...&...&6.27$\pm$0.12&8\\
GQ Lup\tablenotemark{b}&52473&$L'$&...&...&6.39$\pm$0.22&8\\
GQ Lup\tablenotemark{b}&52473&$CH4S$&...&...&6.06$\pm$0.26&8\\
GQ Lup\tablenotemark{a}&53182&$K_s$&734.7$\pm$3&275.48$\pm$0.25&...&9\\
GQ Lup\tablenotemark{b}&53182&$K_s$&...&...&6.00$\pm$0.10&7\\
GQ Lup&53518&$K_s$&735.1$\pm$3&276.00$\pm$0.34&6.39$\pm$0.12&9\\
GQ Lup&53591&$K_s$&733.3$\pm$4&275.87$\pm$0.37&6.3$\pm$0.14&9\\
GQ Lup&53789&$K_s$&729.8$\pm$3&276.14$\pm$0.35&6.29$\pm$0.10&9\\
GQ Lup&53876&$K_s$&731.4$\pm$4&276.06$\pm$0.38&6.07$\pm$0.14&9\\
GQ Lup&53933&$K_s$&733.2$\pm$5&276.26$\pm$0.68&6.43$\pm$0.38&9\\
GQ Lup&54151&$K_s$&730.0$\pm$6&276.04$\pm$0.63&5.79$\pm$0.12&9\\
GSC 06214-00210&54258&$K'$&2203$\pm$2&176.04$\pm$0.06&5.74$\pm$0.05&2\\
GSC 06214-00210&54635&$K'$&2205$\pm$2&175.99$\pm$0.05&5.78$\pm$0.05&2\\
GSC 06214-00210&54635&$J$&2205$\pm$2&176.00$\pm$0.09&6.3$\pm$0.05&2\\
GSC 06214-00210&54983&$K'$&2204$\pm$2&175.91$\pm$0.05&5.88$\pm$0.05&2\\
GSC 06214-00210&55313&$K'$&2206$\pm$2&175.93$\pm$0.05&5.73$\pm$0.05&2\\
GSC 06214-00210&55313&$H$&2203$\pm$2&175.91$\pm$0.05&6.21$\pm$0.05&2\\
GSC 06214-00210\tablenotemark{b}&55313&$L'$&...&...&4.75$\pm$0.05&2\\
GSC 06793-00994&52836&$K_s$&5366$\pm$30&356.1$\pm$0.5&7.76$\pm$0.12&10\\
GSC 06793-00994&54251&$K_s$&5462$\pm$17&357.4$\pm$0.2&7.79$\pm$0.05&2\\
GSC 06794-00156&54251&$K_s$&5973$\pm$18&338.7$\pm$0.2&9.43$\pm$0.05&2\\
PDS 70&53574&$K$&2155$\pm$2&6.3$\pm$0.5&4.68$\pm$0.05&11\\
ROXs 12&52093&$K$&1747$\pm$30&10.3$\pm$0.9&5.75$\pm$0.11&4\\
ROXs 42B&52092&$K$&1137$\pm$30&268.0$\pm$1.5&6.75$\pm$0.40&4\\
ScoPMS 214&52517&$Ks$&3070$\pm$10&121.17$\pm$0.23&5.96$\pm$0.09&10\\
UScoJ1610-1913\tablenotemark{a}&51297&$H$&5872$\pm$70&112.6$\pm$0.7&...&13\\
UScoJ1610-1913\tablenotemark{a}&51297&$J$&5762$\pm$70&114.1$\pm$0.7&...&13\\
UScoJ1610-1913\tablenotemark{a}&51297&$K$&5965$\pm$70&113.6$\pm$0.7&...&13\\
UScoJ1610-1913&54253&$Ks$&5820$\pm$9&114.01$\pm$0.10&3.83$\pm$0.05&14\\

\enddata
\tablenotetext{a}{This observation is only for astrometry; the photometric measurement was too uncertain to be reliable, was reported elsewhere, or was not reported.}
\tablenotetext{b}{This observation is only for photometry; the astrometric measurement was too uncertain to be reliable, was reported elsewhere, or was not reported.}
\tablerefs{
1) \citet[][]{Lafreniere:2010gh},
2) \citet[][]{Ireland:2011fj},
3) \citet[][]{Itoh:2005bs},
4) \citet[][]{Ratzka:2005nx},
5) \citet[][]{White:2001jf},
6) \citet[][]{Janson:2006qf},
7) \citet[][]{Neuhauser:2005ea},
8) \citet[][]{Marois:2007pd},
9) \citet[][]{Neuhauser:2008dp},
10) \citet[][]{Metchev:2009hh},
11) \citet[][]{Riaud:2006hc},
12) \citet[][]{Kohler:2000lo},
13) \citep[][]{Kraus:2009uq},
14) \citep[][]{Kraus:2008zr}.
}
 \end{deluxetable*}

\clearpage

  \begin{deluxetable*}{lrcrrrrl}
 \tabletypesize{\tiny}
 \tablewidth{0pt}
 \tablecaption{New Observations of Known and Candidate Planetary-Mass Companions}
 \tablehead{
 \colhead{Name} & \colhead{JD} & \colhead{Filter} & \colhead{$t_{int}$} & \colhead{$\rho$} & \colhead{PA} & \colhead{$\Delta m$} & \colhead{Tel/Inst}
 \\
 \colhead{} & \colhead{(-2400000)} & \colhead{} & \colhead{(sec)} & \colhead{(mas)} & \colhead{(deg)} & \colhead{(mag)}
 }
 \startdata

1RXS J1609&55736&$K'$&80&2211$\pm$2&27.61$\pm$0.08&7.43$\pm$0.04&Keck-NIRC2\\
2M0415+2818&55157&$K'$&50&2620$\pm$3&89.13$\pm$0.04&5.67$\pm$0.02&Keck-NIRC2\\
DH Tau&54824&$J$&60&2354$\pm$2&139.38$\pm$0.05&5.90$\pm$0.02&Keck-NIRC2\\
DH Tau&54824&$K'$&40&2352$\pm$2&139.35$\pm$0.05&5.92$\pm$0.02&Keck-NIRC2\\
DH Tau&54824&$H$&40&2354$\pm$2&139.39$\pm$0.05&5.97$\pm$0.05&Keck-NIRC2\\
DH Tau&56310&$L'$&160&2350$\pm$2&139.39$\pm$0.05&5.82$\pm$0.02&Keck-NIRC2\\
DoAr 22&56022&$K'$&70&2480$\pm$2&265.32$\pm$0.05&5.97$\pm$0.02&Keck-NIRC2\\
DoAr 22&55735&$K'$&110&2469$\pm$2&264.92$\pm$0.05&5.94$\pm$0.01&Keck-NIRC2\\
DoAr 22&55735&$J$&36&2468$\pm$2&264.93$\pm$0.05&6.22$\pm$0.01&Keck-NIRC2\\
FW Tau&54824&$K'$&40&2282$\pm$2&295.53$\pm$0.08&5.91$\pm$0.03&Keck-NIRC2\\
FW Tau&54824&$H$&80&2285$\pm$2&295.48$\pm$0.08&6.57$\pm$0.02&Keck-NIRC2\\
FW Tau&54824&$J$&120&2280$\pm$2&295.34$\pm$0.17&7.00$\pm$0.04&Keck-NIRC2\\
FW Tau&55849&$K'$&100&2278$\pm$2&295.56$\pm$0.05&5.94$\pm$0.03&Keck-NIRC2\\
FW Tau&56152&$L'$&140&2273$\pm$5&295.50$\pm$0.07&5.15$\pm$0.06&Keck-NIRC2\\
GQ Lup&56023&$K'$&40&722.7$\pm$1.2&277.25$\pm$0.16&6.32$\pm$0.01&Keck-NIRC2\\
GQ Lup&56023&$H$&50&717.3$\pm$1.2&277.37$\pm$0.16&6.09$\pm$0.04&Keck-NIRC2\\
GQ Lup&56023&$L'$&30&713.0$\pm$2.5&277.61$\pm$0.42&6.21$\pm$0.05&Keck-NIRC2\\
GSC 06191-00019&54251&$K_s$&57&3111$\pm$10&321.73$\pm$0.18&8.41$\pm$0.03&Palomar-PHARO\\
GSC 06191-00019&55392&$K'$&120&3203$\pm$3&320.63$\pm$0.06&8.42$\pm$0.07&Keck-NIRC2\\
GSC 06191-00019&56054&$K'$&91&3255$\pm$3&319.93$\pm$0.06&8.49$\pm$0.09&Keck-NIRC2\\
GSC 06214-00210&55716&$K'$&70&2206$\pm$2&175.86$\pm$0.05&5.86$\pm$0.03&Keck-NIRC2\\
GSC 06793-00994&56054&$K'$&54&5537$\pm$6&358.09$\pm$0.02&7.82$\pm$0.05&Keck-NIRC2\\
GSC 06794-00156&56054&$K'$+corona600&40&6009$\pm$6&339.38$\pm$0.02&8.91$\pm$0.02&Keck-NIRC2\\
HD 27659&55157&$K_c$&27&1283$\pm$9&27.44$\pm$0.37&6.71$\pm$0.13&Keck-NIRC2\\
HD 27659&55849&$K'$&500&1304.6$\pm$1.3&27.63$\pm$0.09&7.10$\pm$0.10&Keck-NIRC2\\
HD 27659&55882&$K'$&30&1296$\pm$2&27.81$\pm$0.09&6.87$\pm$0.05&Keck-NIRC2\\
HD 27659&55882&$J_c$&20&1294$\pm$5&27.91$\pm$0.09&7.39$\pm$0.11&Keck-NIRC2\\
HD 27659&55882&$K_c$&90&1301.8$\pm$1.6&27.89$\pm$0.09&6.82$\pm$0.08&Keck-NIRC2\\
HD 27659&55882&$H_c$&131&1298.8$\pm$1.3&27.99$\pm$0.09&6.74$\pm$0.01&Keck-NIRC2\\
HD 27659&55882&$J_c$&90&1299.3$\pm$1.3&28.20$\pm$0.09&6.93$\pm$0.02&Keck-NIRC2\\
HD 27659&55882&$L'$&80&1300$\pm$2&28.09$\pm$0.09&6.73$\pm$0.07&Keck-NIRC2\\
PDS 70&56022&$K'$&60&2321$\pm$2&9.98$\pm$0.05&4.64$\pm$0.01&Keck-NIRC2\\
PDS 70&56022&$H$&60&2321$\pm$2&10.00$\pm$0.05&4.35$\pm$0.03&Keck-NIRC2\\
PDS 70&56022&$J$&60&2321$\pm$2&10.01$\pm$0.05&4.32$\pm$0.02&Keck-NIRC2\\
PDS 70&56022&$L'$&90&2321$\pm$2&9.95$\pm$0.05&5.27$\pm$0.04&Keck-NIRC2\\
ROXs 12&55735&$K'$&100&1781$\pm$2&9.08$\pm$0.06&5.06$\pm$0.01&Keck-NIRC2\\
ROXs 12&55735&$H$&40&1780$\pm$2&9.10$\pm$0.06&5.44$\pm$0.01&Keck-NIRC2\\
ROXs 12&55735&$J$&50&1782$\pm$2&9.08$\pm$0.06&5.62$\pm$0.02&Keck-NIRC2\\
ROXs 12&55735&$L'$&91&1778$\pm$2&9.06$\pm$0.06&4.12$\pm$0.01&Keck-NIRC2\\
ROXs 12&56022&$K'$&70&1783.0$\pm$1.8&8.85$\pm$0.06&5.02$\pm$0.01&Keck-NIRC2\\
ROXs 42B (b)&55735&$K'$&108&1170.5$\pm$1.2&269.98$\pm$0.10&6.38$\pm$0.01&Keck-NIRC2\\
ROXs 42B (b)&55735&$H$&100&1170.2$\pm$1.2&270.02$\pm$0.10&6.89$\pm$0.02&Keck-NIRC2\\
ROXs 42B (b)&55735&$J$&150&1166$\pm$2&269.74$\pm$0.16&7.09$\pm$0.05&Keck-NIRC2\\
ROXs 42B (b)&55735&$L'$&126&1172.1$\pm$1.2&269.99$\pm$0.10&5.70$\pm$0.04&Keck-NIRC2\\
ROXs 42B (b)&56022&$K'$&260&1172.0$\pm$1.2&270.03$\pm$0.10&6.41$\pm$0.01&Keck-NIRC2\\
ROXs 42B (b)&56022&$H$&80&1170.1$\pm$1.2&269.99$\pm$0.10&6.84$\pm$0.05&Keck-NIRC2\\
ROXs 42B (b)&56022&$J$&40&1165.9$\pm$1.7&269.47$\pm$0.10&6.98$\pm$0.12&Keck-NIRC2\\
ROXs 42B (b)&56022&$K_c$&40&1171$\pm$2&269.98$\pm$0.10&6.35$\pm$0.05&Keck-NIRC2\\
ROXs 42B (b)&56511&$K'$&40&1172.5$\pm$1.2&270.25$\pm$0.10&6.32$\pm$0.02&Keck-NIRC2\\
ROXs 42B (cc1)&55735&$K'$&108&575.7$\pm$0.6&223.92$\pm$0.20&6.59$\pm$0.02&Keck-NIRC2\\
ROXs 42B (cc1)&55735&$H$&50&560$\pm$8&224.06$\pm$0.25&6.20$\pm$0.03&Keck-NIRC2\\
ROXs 42B (cc1)&55735&$L'$&126&572$\pm$4&225.0$\pm$0.5&6.59$\pm$0.09&Keck-NIRC2\\
ROXs 42B (cc1)&56022&$K'$&260&568.4$\pm$0.6&225.20$\pm$0.20&6.76$\pm$0.03&Keck-NIRC2\\
ROXs 42B (cc1)&56511&$K'$&40&539.8$\pm$1.2&2255.96$\pm$0.15&6.93$\pm$0.04&Keck-NIRC2\\
ScoPMS 214&54636&$K'$&60&3079$\pm$3&121.13$\pm$0.04&5.78$\pm$0.02&Keck-NIRC2\\
ScoPMS 214&56022&$K'$&91&3071$\pm$3&120.38$\pm$0.04&5.77$\pm$0.05&Keck-NIRC2\\
ScoPMS 214&56032&$K'$&54&3071$\pm$3&120.49$\pm$0.04&5.84$\pm$0.01&Keck-NIRC2\\
ScoPMS 42b (cc1)&54256&Br$\gamma$&44&2980$\pm$3&105.41$\pm$0.05&7.16$\pm$0.13&Keck-NIRC2\\
ScoPMS 42b (cc1)&55392&$K_c$&240&2984$\pm$3&103.51$\pm$0.04&7.41$\pm$0.05&Keck-NIRC2\\
ScoPMS 42b (cc2)&54256&Br$\gamma$&44&4309$\pm$4&332.62$\pm$0.03&6.79$\pm$0.06&Keck-NIRC2\\
ScoPMS 42b (cc2)&55392&$K_c$&240&4362$\pm$4&332.66$\pm$0.03&6.76$\pm$0.02&Keck-NIRC2\\
ScoPMS 42b (cc3)&54256&Br$\gamma$&44&4418$\pm$4&331.77$\pm$0.03&7.22$\pm$0.14&Keck-NIRC2\\
ScoPMS 42b (cc3)&55392&$K_c$&240&4467$\pm$4&331.89$\pm$0.03&7.16$\pm$0.04&Keck-NIRC2\\
UScoJ1608-1935&54257&Br$\gamma$&32&4209$\pm$6&251.36$\pm$0.13&7.19$\pm$0.07&Keck-NIRC2\\
UScoJ1608-1935&56054&$K'$&60&4156$\pm$4&252.45$\pm$0.04&7.35$\pm$0.06&Keck-NIRC2\\
UScoJ1608-1935&56054&$K'$+corona600&40&4160$\pm$4&252.37$\pm$0.03&7.33$\pm$0.10&Keck-NIRC2\\
UScoJ1610-1913&55735&$K'$&80&5836$\pm$6&114.00$\pm$0.02&3.80$\pm$0.01&Keck-NIRC2\\
UScoJ1610-1913&55735&$J$&40&5835$\pm$6&113.98$\pm$0.02&3.86$\pm$0.02&Keck-NIRC2\\
UScoJ1610-1913&55735&$H$&40&5836$\pm$6&113.99$\pm$0.02&4.02$\pm$0.01&Keck-NIRC2\\
UScoJ1610-1913&55735&$L'$&80&5835$\pm$6&114.01$\pm$0.02&3.39$\pm$0.01&Keck-NIRC2\\
UScoJ1610-1913&56022&$K'$&60&5837$\pm$6&113.94$\pm$0.02&3.80$\pm$0.01&Keck-NIRC2\\

 \enddata
  \end{deluxetable*}


  \begin{deluxetable*}{lrrrrrrrrl}
 \tabletypesize{\tiny}
 \tablewidth{0pt}
 \tablecaption{Candidate Companion Properties}
 \tablehead{
 \colhead{Name} & \colhead{$M_{K'}$} &  \colhead{$m_J$} & \colhead{$m_H$} & \colhead{$m_{K'}$} & \colhead{$m_{L'}$}  &
 \colhead{$\mu_{\rho}$} & \colhead{$\mu_{PA}$} &  \colhead{$\rho$} & \colhead{$M$} \\
\colhead{}  & \colhead{(mag)} & \colhead{(mag)} & \colhead{(mag)} & \colhead{(mag)} & \colhead{(mag)} &
 \colhead{(mas/yr)} & \colhead{(mas/yr)} &  \colhead{(AU)} & \colhead{($M_{Jup}$)}
 }
 \startdata

\multicolumn{5}{l}{\it New Companions}\\
FW Tau&9.51$\pm$0.24&17.34$\pm$0.07&16.25$\pm$0.07&15.32$\pm$0.07&14.25$\pm$0.10&-1.2$\pm$0.2&0.8$\pm$0.7&330$\pm$30&$10 \pm 4$\\
ROXs 12&8.93$\pm$0.19&16.64$\pm$0.07&15.37$\pm$0.08&14.32$\pm$0.06&12.55$\pm$0.09&3.4$\pm$2.0&-6.5$\pm$1.9&210$\pm$20&$16 \pm 4$\\
ROXs 42B (b)&9.62$\pm$0.19&16.99$\pm$0.07&15.88$\pm$0.06&15.01$\pm$0.05&14.15$\pm$0.09&0.2$\pm$1.1&-0.7$\pm$1.6&140$\pm$10&$10 \pm 4$\\

\multicolumn{5}{l}{\it Known Companions}\\
DH Tau&8.34$\pm$0.24&15.69$\pm$0.07&14.88$\pm$0.07&14.15$\pm$0.07&12.93$\pm$0.08&0.2$\pm$0.2&-1.0$\pm$0.3&340$\pm$30&$18 \pm 4$\\
GQ Lup&7.37$\pm$0.22&14.82$\pm$0.11&13.79$\pm$0.08&13.33$\pm$0.06&12.29$\pm$0.10&-2.0$\pm$0.2&2.5$\pm$0.3&110$\pm$10&$31 \pm 3$\\
1RXS J1609&10.42$\pm$0.23&17.90$\pm$0.13&16.87$\pm$0.09&16.23$\pm$0.06&14.87$\pm$0.31&-0.5$\pm$0.9&-1.0$\pm$1.1&320$\pm$30&$10 \pm 2$\\
UScoJ1610-1913&7.00$\pm$0.23&13.89$\pm$0.08&13.28$\pm$0.07&12.80$\pm$0.06&12.09$\pm$0.09&1.8$\pm$1.8&-1.2$\pm$1.6&850$\pm$90&$70 \pm 10$\\
GSC 6214-210&9.15$\pm$0.23&16.30$\pm$0.08&15.55$\pm$0.07&14.95$\pm$0.06&13.83$\pm$0.09&0.4$\pm$0.6&-1.7$\pm$0.6&320$\pm$30&$16 \pm 1$\\

\multicolumn{5}{l}{\it Unassociated Objects}\\
2M04153916&...&...&...&14.91$\pm$0.07&...&...&...&&\\
HD 27659&...&14.31$\pm$0.07&14.03$\pm$0.07&14.09$\pm$0.07&14.37$\pm$0.10&-5.5$\pm$4.2&7.8$\pm$4.2&&\\
PDS 70&...&13.87$\pm$0.07&13.17$\pm$0.08&13.20$\pm$0.07&13.30$\pm$0.09&24.7$\pm$0.3&26.0$\pm$3.0&&\\
GSC 06191-00019&...&...&...&16.54$\pm$0.07&...&29.0$\pm$1.7&-21.2$\pm$1.9&&\\
UScoJ160823.8-193551&...&...&...&16.53$\pm$0.07&...&-10.4$\pm$1.4&15.4$\pm$2.0&&\\
ScoPMS 42b (cc1)&...&...&...&17.00$\pm$0.07&...&1.3$\pm$1.4&-31.9$\pm$1.1&&\\
ScoPMS 42b (cc2)&...&...&...&16.39$\pm$0.07&...&17.0$\pm$1.8&0.9$\pm$1.1&&\\
ScoPMS 42b (cc3)&...&...&...&16.79$\pm$0.07&...&15.8$\pm$1.8&2.9$\pm$1.1&&\\
GSC 06793-00994&...&...&...&16.41$\pm$0.06&...&17.4$\pm$2.6&16.2$\pm$3.2&&\\
GSC 06794-00156&...&...&...&16.51$\pm$0.07&...&7.3$\pm$3.8&14.6$\pm$4.3&&\\
DoAr 22&...&15.95$\pm$0.07&...&14.96$\pm$0.06&...&16.0$\pm$2.1&24.1$\pm$2.2&&\\
ScoPMS 214&...&...&...&13.57$\pm$0.06&...&-1.1$\pm$0.8&-8.6$\pm$0.6&&\\
ROXs 42B (cc1)&...&...&15.22$\pm$0.07&15.46$\pm$0.06&15.04$\pm$0.12&-9.1$\pm$1.1&14.4$\pm$3.2&&\\
 
 \enddata
\tablecomments{Each photometry measurement reflects the weighted mean of all measurements in Tables 2 and 3, except those which are footnoted as having been explicitly excluded. The minimum uncertainty at an epoch is assumed to be $\sigma = 0.05$ mag (due to stellar variability of the primary). The uncertainties here also reflect the uncertainty in the primary star magnitudes taken from 2MASS and WISE, as well as the color-SpT relation used to convert WISE $W1$ to MKO $L'$. Stellar variability at those epochs also imposes a systematic uncertainty of $\sigma = 0.05$ mag. The magnitudes reported here have not been corrected for extinction, but we have de-extincted or dereddened using the $A_V$ inferred for each corresponding primary (Table 1) before computing masses or colors (e.g., Figures 7--10). The uncertainties in projected separations are driven entirely by the uncertainty in distance, and hence have the same fractional uncertainty as the quoted values (Section 2.3).}
 \end{deluxetable*}

\end{document}